\documentclass[aps,showpacs,superscriptaddress,twocolumn]{revtex4}

\usepackage{amsfonts}
\usepackage{amsmath}
\usepackage{amsthm}
\usepackage{amscd}
\usepackage{amssymb}
\usepackage{amsxtra}
\usepackage{bm}           
\usepackage{bbm}
\usepackage{exscale}
\usepackage{epic,rotating}
\usepackage{graphics,color,dsfont}
\usepackage{latexsym}
\usepackage{enumerate}
\usepackage{dcolumn}      
\usepackage{pstricks,pst-grad,fancybox,graphics}
\usepackage[scanall]{psfrag}
\usepackage[dvips]{epsfig}

\newcommand{\Kommut}[2]{\ensuremath{[ #1,#2]}}
\newcommand{\Akommut}[2]{\ensuremath{\{ #1,#2 \}}}

\newcommand{\gh}{\ensuremath{\hat{G}}}
\newcommand{\C}{\ensuremath{\mathbbm C}}
\newcommand{\R}{\ensuremath{\mathbbm R}}

\newcommand{\bra}[1]{\ensuremath{\langle#1|}}

\newcommand{\ket}[1]{\ensuremath{|#1\rangle}}

\newcommand{\ketbra}[1]{\ensuremath{| #1 \rangle \langle #1 |}}

\newcommand{\Eins}{\ensuremath{\mathbbm 1}}
\newcommand{\eins}{\ensuremath{\mathbbm 1}}

\newcommand{\HH}{\ensuremath{\mathcal{H}}}

\newcommand{\fg}{\ensuremath{\mathfrak{g}}}
\newcommand{\fk}{\ensuremath{\mathfrak{k}}}

\newcommand{\fC}{\ensuremath{\mathfrak{C}}}
\newcommand{\fB}{\ensuremath{\mathfrak{B}}}
\newcommand{\fA}{\ensuremath{\mathfrak{A}}}

\newcommand{\bear}{\begin{eqnarray}}
\newcommand{\eear}{\end{eqnarray}}
\newcommand{\bearn}{\begin{eqnarray*}}
\newcommand{\eearn}{\end{eqnarray*}}
\newcommand{\kommentar}[1]{}

\newcommand{\mean}[1]{\ensuremath{\langle #1 \rangle}}

\newcommand{\tr}{{\rm tr}}

\newcommand{\bc}{\begin{center}}
\newcommand{\ec}{\end{center}}

\newcommand{\zz}{\mathbbm{Z}}

\newcommand{\NORM}{\ensuremath{|\!|\!|}}

\renewcommand{\qed}{\ensuremath{\hfill \blacksquare}\medskip}

\newtheorem{thm}{Theorem}[section]
\newtheorem{lem}[thm]{Lemma}
\newtheorem{corol}[thm]{Corollary}
\newtheorem{proposition}[thm]{Proposition}
\newtheorem{defn}[thm]{Definition}
\newtheorem{rem}[thm]{Remark}
\renewcommand{\vr}{\ensuremath{\varrho}}
\def\bi#1\ei {\begin{itemize}#1\end{itemize}}
\def\bea#1\eea {\begin{align}#1\end{align}}
\def\bean#1\eean {\begin{align*}#1\end{align*}}
\def\ben#1\een {\begin{equation*}#1\end{equation*}}
\def\be#1\ee {\begin{equation}#1\end{equation}}
\def\bes#1\ees {\begin{equation}\begin{split}#1\end{split}\end{equation}}

\newcommand{\halbe}{\frac{1}{2}}

\newcommand{\sz}[1]{\sigma^z_{#1}}
\newcommand{\sx}[1]{\sigma^x_{#1}}
\newcommand{\sy}[1]{\sigma^y_{#1}}

\begin{document}

\title{Unifying several separability conditions using the covariance matrix criterion}

\author{O.~Gittsovich}
\affiliation{Institut f\"ur Theoretische Physik,
Universit\"at Innsbruck, Technikerstra{\ss}e 25,
6020 Innsbruck, Austria}

\affiliation{Institut f\"ur Quantenoptik und Quanteninformation,
~\"Osterreichische Akademie der Wissenschaften, Otto-Hittmair-Platz 1,
6020 Innsbruck, Austria}

\author{O.~G\"uhne}
\affiliation{Institut f\"ur Quantenoptik und Quanteninformation,
~\"Osterreichische Akademie der Wissenschaften, Otto-Hittmair-Platz 1,
6020 Innsbruck, Austria}

\affiliation{Institut f\"ur Theoretische Physik,
Universit\"at Innsbruck, Technikerstra{\ss}e 25,
6020 Innsbruck, Austria}

\author{P.~Hyllus}
\affiliation{Institut f\"ur Theoretische Physik, Universit\"at
Hannover, Appelstra{\ss}e 2, 30167 Hannover, Germany}

\author{J.~Eisert}
\affiliation{QOLS, Blackett Laboratory, Imperial College London,
Prince Consort Road, London SW7 2BW, UK}

\affiliation{Institute for Mathematical Sciences,
Imperial College London,
Prince's Gate, London SW7 2PE, UK}

\begin{abstract}
We present a framework for deciding whether a quantum state is separable
or entangled using covariance
matrices of locally measurable observables.
This leads to the covariance matrix criterion as a general separability
criterion. We demonstrate that this criterion allows to detect many states
where the familiar criterion of the positivity of the partial transpose
fails.
It turns out that a large number of criteria which have been proposed
to complement the positive partial transpose criterion -- the computable
cross norm or realignment criterion, the criterion based on local uncertainty
relations, criteria derived from extensions of the realignment map, and others --
are in fact a corollary of the covariance matrix criterion.

\end{abstract}


\date{\today}

\maketitle

\section{Introduction}

Entanglement is the feature of quantum theory that renders it
crucially different from a classical statistical theory. It also plays the
central role in quantum information science, as a resource for
information processing tasks. Consequently, a lot of effort has been
made in the last decade to understand the meaning and the structure
of entangled states \cite{hororeview,morereviews}.
One of the most elementary yet
notorious questions is how to find good criteria to decide whether a state
is entangled or classically correlated in the first place. Formally,
one asks whether a given state $\rho$ in a bipartite system is
contained in the
convex hull of product states and can hence be written as
\be
\rho = \sum_k p_k \ketbra{a_k}\otimes \ketbra{b_k},
\ee
where the coefficients $p_k$ form a probability distribution. If so, all
correlations can come from classical shared randomness, and a state is
called classically correlated or separable. Otherwise, $\rho$ is an entangled
state.
The decision problem of deciding whether a state is entangled or separable
is known to be a computationally hard problem in the physical
dimension, for a certain scaling of the error in the weak membership
problem \cite{Gurvits}. Yet, the problem one typically faces is the one
where one has just a physical state given -- having some fixed dimension -- and one aims
at finding criteria to make a judgment based on these criteria. Not
surprisingly, given the central status of entanglement in quantum
information theory, a lot of effort has been devoted to identifying
such good and practical and  computable criteria for
separability in composite quantum states.

Historically, the first criterion of this sort was derived from
the observation that every separable state will have a positive
partial transpose, and that the positivity of the latter can hence be
used as an entanglement criterion {\it (PPT criterion)} \cite{ppt1,ppt2}.
This criterion later turned out to be necessary and sufficient for separability for
low dimensional systems ($2\times 2$ and $2\times 3$), whereas
in higher dimensions this is no longer the case
\cite{horobound}. The PPT criterion is an example of a {\it criterion based
on positive maps}: In fact, it has been proven that a state
$\rho$ is separable if and only if for any  positive map $\Lambda$ the
operator $(\Eins\otimes\Lambda)(\rho)$ is also  positive \cite{ppt2}.
Consequently, the systematic
investigation of positive maps has led to a number of new
separability criteria \cite{positivemaps}. A quite remarkable criterion
of this type is the {\it reduction criterion}, which is equivalent to the PPT
criterion for $2\times 2$ and $2\times 3$ cases and weaker for higher
dimensions
\cite{redcrit}. There are also other criteria that turned out to be directly
related to the PPT criterion: The {\it majorization criterion}
\cite{kempenielsen}
and also {\it entropic criteria} \cite{entropies} have been shown to be weaker
than the PPT criterion \cite{nagasaki, vollbrecht}. Moreover,
one can extend the PPT condition to a test based on a complete
hierarchy of symmetric extensions, where each step constitutes
a {\it semidefinite program} \cite{doherty}
(another complete family of semidefinite tests has been
described in \cite{ourold}).
In such a hierarchy, every entangled state is necessarily
detected as such in some step of the hierarchy.
However, the steps in the hierarchy require more and more computational effort,
which makes this approach difficult already for modest system sizes.

Apart from these criteria that are directly related to the PPT criterion, a
number
of other separability criteria have been suggested where such a connection
seemingly does not exist.  The most prominent criterion of this type is the
 {\it computable cross-norm}  or
{\it realignment criterion} (CCNR) \cite{ccncrit}. Other criteria in this category
 use the
{\it Bloch representation} of density matrices \cite{deVic, deVicquant}, {\it local
uncertainty relations}
(LURs) \cite{LUR}, {\it local orthogonal observables}
\cite{YuLiu}, or {\it
 extensions of the
realignment map} \cite{china1,china2}. These criteria are
complementing the PPT
criterion in an interesting way: In fact they detect some states as being entangled
where the PPT criterion fails. They also do not rely on positive maps.

At first sight, one might think that these criteria form a collection of
quite beautiful, but strangely disconnected results. They have been derived
using a variety of unrelated methods, and their connection often seems
quite unclear. It is the main purpose of this work
to develop a framework for the systematic
understanding of all these latter approaches.

In Ref.~\cite{wir} we have proposed to investigate the separability problem
using covariance matrices (CMs) of certain observables. In this
context we have developed a
separability criterion in terms of covariance matrices
({\it covariance matrix criterion}
or CMC). We have shown that the CMC is, when augmented
with appropriate local filtering,  despite its simplicity a surprisingly
strong entanglement criterion, which can detect states where
the PPT criterion fails
and which is at the same time necessary and
sufficient for two qubits. Here we complete this approach
and present new results in
various directions. Specifically, we show that a number of
criteria which have been proposed to improve the PPT
criterion -- namely the
CCNR criterion \cite{ccncrit}, a criterion
using the Bloch representation \cite{deVic, deVicquant}, the LURS \cite{LUR}, and recent
criteria
from Refs.\  \cite{china1,china2} -- follow directly from the CMC. In
this
way the CMC can be seen as complementary to the PPT criterion. We also
tighten
previous formulations of the CMC. We discuss several examples, and compare
the
performance of the criteria to instances of random states from a families of
bound entangled states.

This manuscript is organized as follows: In the second section,
we introduce CMs and discuss their mathematical properties.
Those readers who are mainly interested in separability criteria
may only consume Definitions \ref{cmdefinition} and \ref{blockcmdefinition}
and Propositions \ref{CMform}, \ref{CMformsym} and \ref{gammaprop}
and may then directly jump to Section III.
In that Section, we introduce the CMC
and evaluate it in several different ways. By doing this we establish the
mentioned connection to the other separability criteria which will turn
out to be corollaries of the former. In Section IV of the paper we will
consider the  connection between the CMC and the LURs. In the
fifth section we will scrutinize the CMC
for the two qubit case. In Section VI we will
assess the strength  of the mentioned
criteria by considering a family of bound entangled states. We
will then conclude and elaborate on
possible extensions of the work presented here. Some
more technical
proofs of our theorems will finally be presented in the
Appendix.

\section{Covariance matrices}
\label{cmprop}

In this section we will investigate covariance matrices as our main tool.
In the first subsection we will introduce the  different
definitions of CMs \cite{wir,china2,mira}
and fix our notation. In the second subsection we will address the
question to which extent CMs can be used as a unique description of quantum
states besides density matrices. Finally, in the third and fourth subsection we will mention
and prove some useful properties of CMs, which will be used later in
our study of entanglement.

\subsection{Definition of covariance matrices}

In what follows let $\rho$ be a pure or mixed quantum state, described by a
 (positive) density operator in a $d$-dimensional Hilbert space
$\mathcal{H}$ and let $\{M_k:k=1,\dots, N\}$ a suitable set of observables.
Unless
stated otherwise, we will always assume that these observables are orthonormal
observables with respect to the Hilbert-Schmidt scalar product between
observables, i.e., they fulfill
\be
    \tr(M_i M_j) = \delta_{i,j}.
\ee
Furthermore, we will typically assume that the $M_i$ form a complete basis
and span the whole observable algebra. This implies that there are $N=d^2$
different $M_i,$ and that any other observable can be
expressed as a linear combination of the $M_i.$

As an example for such a set of observables for the case
of a single qubit,
 one can consider the (appropriately normalized)
 Pauli
 matrices,
\be
\label{twoqubitloos}
M_1= \frac{\eins}{\sqrt{2}}, \;\;
M_2 =\frac{\sigma_x}{\sqrt{2}}, \;\;
M_3 =\frac{\sigma_y}{\sqrt{2}}, \;\;
M_4 =\frac{\sigma_z}{\sqrt{2}}.
\ee
We can now formulate the main definitions for this
work.

\begin{defn}[Covariance matrix]
\label{cmdefinition}
The $d^2\times d^2$
covariance matrix $\gamma = \gamma (\rho,\{M_k\})$
and the $d^2\times d^2$
symmetrized covariance matrix $\gamma^S = \gamma^S(\rho,\{M_k\})$
are defined by their matrix entries as
\bear
\gamma_{i,j}&=&\mean{M_iM_j}-\mean{M_i}\mean{M_j},
\label{cmdef}\\
\gamma^{S}_{i,j}&=&\frac{\mean{M_iM_j}+\mean{M_jM_i}}{2}-\mean{M_i}\mean{
M_j}.
\label{cmdefsym}
\eear
Sometimes, the difference between the linear part of a CM
and the nonlinear part becomes relevant. Therefore, we define
the linear part of $\gamma$ as $\fg_{i,j}=\mean{M_iM_j}$
and the linear part of the symmetric CM as
$\fg^S_{i,j}=\mean{M_iM_j+M_j M_i}/2.$
\end{defn}
We will often for simplicity of notation also write $\gamma(\rho)$
or $\gamma(\{M_k\})$
instead of $\gamma(\rho,\{M_k\})$, or simply $\gamma$.
We will also sometimes indicate with respect
to what state an expectation value is taken, so
$\mean{M_i}=\mean{M_i}_\rho$.
It is straightforward to see that $\gamma$ is a complex Hermitian matrix.
The matrix $\gamma^S$ in turn
is real and symmetric. Both $\gamma$ and $\gamma^S$ are
positive semidefinite, $\gamma,\gamma^S\geq 0$ \cite{robertson}.

Note finally that for odd $d$, there is another basis of
orthonormal observables that can equally be used and that
is commonly employed in the mathematical physics literature in
the context of {\it discrete Weyl} systems \cite{Vourdas}.
Let $A(0,0)$ be the
parity operator that maps $P(0,0):|x\rangle\mapsto |-x\rangle$,
where $|x\rangle\in \{|0\rangle,\dots, |d-1\rangle\}$, meant modulo
$d$. Then, for $(q,p)\in \zz_d^2$
let
\begin{equation}
    P(q,p)=W(q,p) P(0,0)W(q,p)^\dagger
\end{equation}
the translated versions of $P(0,0)$ in discrete phase space,
where $W(q,p)$ are the discrete Weyl operators \cite{Weyl}.
The operators
$\{M_{(q,p)}\}=
\{P(q,p)\sqrt{d}\}$ then form a set of Hilbert-Schmidt orthonormal
Hermitian matrices. This is the standard set of observables
when phase-space methods are made use of.

\subsection{Covariance matrices for bipartite systems}

In the focus of this work is the situation where the
Hilbert space is a tensor product
$\HH = \HH^A \otimes \HH^B$ of Hilbert spaces of two subsystems
$A$ and $B$.  We consider finite-dimensional systems,
and denote the dimension of
$\HH^A$ ($\HH^B$) with $d_A$ ($d_B$), respectively, such that
the  dimension of the tensor product Hilbert space
is $d = d_A \times d_B.$ We can choose a basis of
the observable algebra in $A$ as $\{A_k : k=1, \dots ,d_A^2\}$
and in $B$ as $\{B_k : k=1, \dots ,d_B^2\}$,
and consider the set of $d_A^2 + d_B^2$ observables
\begin{equation}
    \{M_k\} = \{ A_k \otimes \eins, \eins \otimes B_k\}.
\end{equation}
Note that this set is not tomographically complete, since
observables like $A_k \otimes B_l$ are missing. However, this
set can be employed to define a very useful form of CMs.

\begin{defn}[Block covariance matrices]
\label{blockcmdefinition}
Let $\rho$ be a state of a bi-partite system, and let
$M_k = \{ A_k \otimes \eins, \eins \otimes B_k\}$
be a set of observables as outlined above.
Then, the block covariance matrix $\gamma(\rho,\{M_k\})$
has the entries
$\gamma_{i,j}=\mean{M_iM_j}-\mean{M_i}\mean{M_j}$
and consequently a block structure:
\be
\gamma
=
\begin{pmatrix}
A & C
\\
C^T & B
\end{pmatrix},
\label{BCM}
\ee
where $A = \gamma(\rho_A, \{A_k\})$ and
$B = \gamma(\rho_B, \{B_k\})$ are CMs of the
reduced states of systems $A$ and $B$,
and
\be
C_{i,j}=\mean{A_i\otimes B_j}-\mean{A_i}\mean{B_j}.
\ee
Similarly, we can define a symmetric
block covariance matrix $\gamma^S(\{M_k\}),$ for which
$A$ and $B$ are the corresponding symmetric CMs, while
$C$ remains unchanged.
\end{defn}

\subsection{Covariance matrices as description of quantum states}
\label{description}

Is it possible to completely reconstruct the state from a given CM?
As our separability criterion uses the CM to decide separability, this
question is important in order to understand, whether all states can
be detected. We will discuss it in this subsection. Let us first show
how CMs depend on the set of observables $\{M_k:k=1,\dots, N\}$:

\begin{proposition}[Transformation of covariance matrices]
\label{bastrans}
Let $\gamma(\{M_k\})$ be a CM as defined in (\ref{cmdef}). If $\{K_k\}$
is another set of observables, connected to the $\{M_k\}$ by a basis
transformation $K_i=\sum_{i=1}^N O_{i,j} M_j$ with some matrix $O$
then  $\gamma(\{K_k\})$ is given by
\be
\gamma(\{K_k\})=O \gamma(\{M_k\}) O^T.
\ee
Note that $O$ is an orthogonal matrix if $K_i$ and $M_i$ are
orthonormal bases.
\end{proposition}
{\it Proof:} A direct calculation gives
\begin{eqnarray}
\gamma(\{K_k\})_{i,j}& =&
\sum_{l,m}\mean{O_{i,l}M_l O_{j,m}M_m}
- \mean{O_{i,l}M_l}\mean{O_{j,m}M_m}\nonumber\\
&=&\sum_{l,m} O_{i ,l}\gamma(\{M_k\})_{l,m}O^T_{m,j},
\end{eqnarray}
which proves the claim.
\qed

The main point is that the previous proposition allows us to choose the
basis
which we want to express our CM in arbitrarily, since we know how the CM
will be transformed under a basis transformation in the space of
observables.

We can now come back to the initial question: Suppose we are given some CM
with a fixed basis of observables.  Are we able to reconstruct the
physical state from this CM uniquely?  We will start answering this
question by considering a single system.

\begin{proposition}[Characterization of states via non-symmetric covariance
matrices]
\label{cm-dm}
Given a non-symmetric CM with tomographically complete set of observables,
we can reconstruct the corresponding physical state unambiguously.
\end{proposition}
{\it Proof:} We choose the following basis of the observables:
\bear
D_i &=& \ketbra{i}, \;\;\; i = 1, \dots ,d, \\
X_{i,j} &=& \frac{1}{\sqrt{2}} (\ket{i}\bra{j}+\ket{j}\bra{i}), \;\;\; 1
\leq  i < j \leq d, \\
Y_{k,l} &=& \frac{i}{\sqrt{2}} (\ket{k}\bra{l}-\ket{l}\bra{k}), \;\;\; 1
\leq  k<l \leq d .
\label{standbas}
\eear
These observables form an orthonormal basis, and we will refer to this
basis as to the {\it standard basis} later on. As in any basis $M_k,$ we can
write the
state as $\rho=\sum_k \mean{M_k} M_k$, it suffices to know the first
moments $\mean{M_k}.$ From Eq.~(\ref{cmdef}) one can see that
$\gamma_{i,j}-\gamma_{j,i}=\mean{\Kommut{M_i}{M_j}}$. In the following we
will show that in the chosen basis, all first moments can be obtained from
expectation values of commutators.

For the chosen standard basis we can explicitly calculate all commutators
\bear
\Kommut{D_k}{X_{k,l}}&=&\frac{i}{\sqrt{2}}Y_{k,l},\;\;\;
\Kommut{D_k}{Y_{k,l}}= -\frac{i}{\sqrt{2}}X_{k,l},\\
\Kommut{X_{k,l}}{Y_{k,l}}&=&i(\ketbra{k}-\ketbra{l}).
\eear
Hence, all expectation values of the $X_{i,j}$ and $Y_{k,l}$ can be
calculated. The same is true for the diagonal elements: Using the fact
that the trace of the density matrix is equal to one, we can calculate all the
diagonal elements from the mean values of $\Kommut{X_{k,l}}{Y_{k,l}}.$
$\qed$

Clearly, the same approach can be used for bipartite systems, if
we use the CM in the full (and not in a block) form. In this case
we can use a product basis $\{\ket{i_1,i_2}\}$. Identifying $(i_1,i_2)
=: i$ we can define the standard basis as above and find all first moments
from the covariance matrix.

As we have seen, the non-symmetric CM defined in Eq.~(\ref{cmdef})
describes the physical state completely. The knowledge of the symmetric
CM in Eq.~(\ref{cmdefsym})
is, however, not enough:

\begin{proposition}[Inequivalence of
states and symmetric covariance
matrices]\label{nonequivalence}
The knowledge of the symmetric CM $\gamma^S$ does,
in general, not determine
the  state $\rho$ completely.
\end{proposition}

{\it Proof:} We prove the claim by providing a counterexample. Let us
take a single qubit. As observables we take the appropriate normalized
Pauli matrices. The symmetric CM has the following entries
\bear
\gamma^S_{0,j}&=&
\frac{\mean{\eins\sigma_j}+\mean{\sigma_j\eins}}{4}-\frac{\mean{\eins}\mean
{\sigma_j}}{2}
= 0 = \gamma^S_{i,0},
\\
\gamma^S_{i,j}&=&
\frac{\mean{\Akommut{\sigma_i}{\sigma_j}}}{4}-\frac{\mean{\sigma_i}\mean{\sigma_j}}{2}
=
\frac{\delta_{i,j}-\mean{\sigma_i}\mean{\sigma_j}}{2}.
\label{onequbitcm}
\eear
{From} this we can
determine the norm of the mean value of the spin component in
a certain direction,
but not its sign. Hence we know the length of the
Bloch vector of the system, up
to some reflection to the origin, which corresponds to
simultaneous change of signs of all $\mean{\sigma_i}$'s.

One might think that the case of one qubit constitutes a special case.
However, the same ambiguity will arise if one embedded
a qubit in a higher dimensional, say, three level system. As it
can be checked, the additional observables in the basis of
observables $\{M_k\}$ will not provide any further information.
$\qed$

To summarize: The knowledge of the symmetric CM of a qubit alone
is not sufficient to decide between two alternatives of states
which have opposite (symmetric to the origin) Bloch vectors.
Also, merely the additional knowledge of a single bit (the sign)
is needed to make this
correspondence unambiguous. This, however, is specific to the
qubit case.
We will now turn to investigating the same question for
the block CM defined in  Eq.~(\ref{BCM}):

\begin{proposition}[Relationship between bipartite states and block covariance
matrices]
For block CMs $\gamma$ and $\gamma^S$ on a bipartite system,
the following statements hold:
\bi
\item[{(i)}] The (non-symmetric) block CM $\gamma$ determines
the bipartite state
$\rho_{AB}$ completely.
\item[{(ii)}] The symmetric block $\gamma^S$ does not determine $\rho_{AB}$
completely.
\ei
\end{proposition}
{\it Proof:} Obviously, given a non-symmetric block CM for the set of
variables $A_k\otimes\eins$ and $\eins\otimes B_l$ we can determine
first all $\mean{A_k}$ and $\mean{B_l}$ for the reduced state $\rho_A$
in the same
way as in Proposition \ref{cm-dm} from the blocks $A$ and
$B$ of $\gamma.$ Then, knowing the block $C$ we can fix the
rest $\mean{A_k\otimes B_l}$ as
\be
\mean{A_k\otimes B_l} = C_{k,l} + \mean{A_k}\mean{B_l}
\label{restbloch}
\ee
and hence {\it (i)} is proved.

The validity of {\it (ii)} is straightforward to see for two qubit
states, as there will be the same lack of information on the mean
values of
observables as in Proposition \ref{nonequivalence} and
hence $\gamma^S_{AB}$
does not provide the whole
information about the state.
$\qed$

The fact that the symmetric block CM $\gamma^S$ does not determine the state
completely will later be important for the discussion of our entanglement
criteria. Therefore, let us investigate this correspondence for the case of
two qubits in some more detail.
For that, let $A_i$ and $B_j$ be Pauli matrices. We may write
the state in the form
\be
\rho_{AB}=\frac{1}{4}
\sum_{i,j}\lambda_{i,j}\sigma^A_{i}\otimes\sigma^B_{j},
\ee
where $\lambda_{i,j}=\tr(\rho\sigma^A_{i}\otimes\sigma^B_{j})$.

As one can see from Eq.~(\ref{restbloch}) we have two possibilities
of changing the $\lambda_{i,j}$ while keeping the $C_{i,j}$ invariant:
We can (i) flip the signs of both of the Bloch vectors of the reduced density
matrices, ($\lambda_{0,j}$ and $\lambda_{i,0}$ for $i,j = 1,2,3)$,
while keeping the left hand side of  Eq.~(\ref{restbloch}) invariant.
Alternatively, we can (ii) flip the sign of only one of them, which implies
that we also have to change the left hand side of  Eq.~(\ref{restbloch}).

Concerning (i), one can directly calculate the
transformed state $\rho^{\text{inv}}$. It turns out that the
eigenvalues of $\rho$ and $\rho^{\text{inv}}$ are the same,
suggesting that they are connected by a unitary
transformation maybe in addition with a global transposition
which transforms one state to the other. Actually the
unitary transformation is a {\it local} unitary one, and
one has the following transformation:
\bes
&\left(\rho^{\text{inv}} \right)^T = U^{\dagger}\rho_{AB}U,\\
&U=\begin{pmatrix} 0 & -1\\1 & 0\end{pmatrix}\otimes
\begin{pmatrix}0 & 1\\-1 & 0\end{pmatrix}
\label{beideBloch}.
\ees
Since there is no physical process which
corresponds to a transposition of a state, there are two physically
different
states $\rho$ and $\rho^{\text{inv}}$ which give rise to
the same covariance matrix and which are
connected by the simultaneous flip of the Bloch vectors of their subsystems.
Nevertheless we can see from Eq.~(\ref{beideBloch}) that these
states have the same entanglement properties, because there is a local unitary
operation in addition to a
global transposition connecting them. These
transformations do not change
the outcome of the PPT criterion, and in fact do not change the
entanglement properties of any two-qubit quantum state.

Concerning (ii), it also possible to flip the Bloch vector of
only one of the subsystems in a such a way that the whole covariance matrix
will remain unaltered. This kind of transformation is done by
\bes
\mean{\sigma^A_{i}} &\mapsto -\mean{\sigma^A_{i}},\\
\mean{\sigma^A_{i}\otimes\sigma^B_{j}} &\mapsto
\mean{\sigma^A_{i}\otimes\sigma^B_{j}}-2\mean{\sigma^A_{i}}
\mean{\sigma^B_{j}},
\ees
resulting in a transformation of $\rho$ to a different $\rho^{\text{inv}}$.
Such a change of the state is nontrivial and can give rise to
a matrix  $\rho^{\text{inv}}$
with negative eigenvalues, which clearly does not correspond to any state.
Two more cases that should be discussed. As one can see from a
numerical search,
there are some states $\rho,$ for which $\rho^{\text{inv}}$  is still a state
and $\rho$ and $\rho^{\text{inv}}$ are either both separable or both entangled.
But there exist also states which alter their separability properties after
a Bloch vector inversion. As an example of states where $\rho$ and
$\rho^{\text{inv}}$ have different separability properties, consider the
states of the form
\be
\rho_{\varepsilon}=\frac{\varepsilon}{2}\begin{pmatrix}
1+r & 0 & 0 & t\\
0   & 0 & 0 & 0\\
0   & 0 & s-r & 0\\
t   & 0 & 0 & 1-s
\end{pmatrix} + (1-\varepsilon)\begin{pmatrix}
0   & 0 & 0 & 0\\
0   & 1 & 0 & 0\\
0   & 0 & 0 & 0\\
0   & 0 & 0 & 0
\end{pmatrix}.
\ee
$\rho_{\varepsilon}$ is a slight modification of the family of the states
introduced
in Ref.\ \cite{Rud2} and which are known to be detected by PPT but not by CCNR
criterion for certain parameters. The inverted form
$\rho_\varepsilon^{\text{inv}}$ of this
states can be calculated analytically.
The states $\rho_{\varepsilon}$ are known to be PPT
for ($t=0,\varepsilon=1$).
Going away from $\varepsilon=1$ and changing other
parameters one can find regions
where $\rho_{\varepsilon}$ and $\rho_{\varepsilon}^{\text{inv}}$
have different entanglement
properties.

\begin{figure}[t]
\begin{center}
\includegraphics[width=0.95\columnwidth]{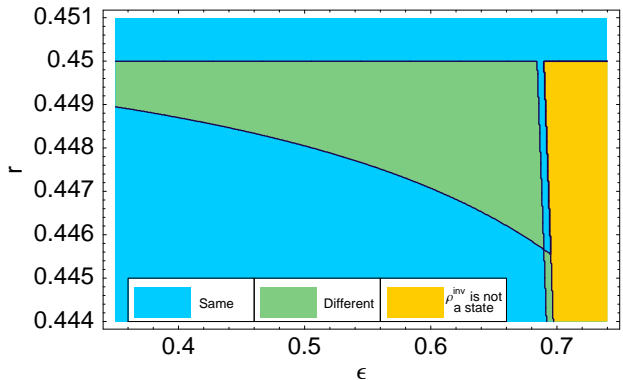}
\caption{(Color online) Entanglement properties of $\rho_{\varepsilon}$ and
$\rho^{\text{inv}}_{\varepsilon}$ are revealed by the PPT criterion
for $s=0.45$ and  $t=\frac{1}{16}$. $\varepsilon$ and $r$
are varied. Three regions corresponds to three different
cases. The region ``Same" corresponds to the case
where $\rho_{\varepsilon}$ and $\rho^{\text{inv}}_{\varepsilon}$
are either both separable or both entangled. The region
``Different" corresponds to the situation where $\rho_{\varepsilon}$
is separable but $\rho^{\text{inv}}_{\varepsilon}$ is entangled or
{\it vice versa}. The last region consists of states
$\rho_{\varepsilon}$ for which the inversion of the Bloch vector
of one of the subsystems leads to $\rho^{\text{inv}}_{\varepsilon}$
which is not positive semi-definite anymore.}
\label{lambdas}
\end{center}
\end{figure}

As we can see from Fig.~\ref{lambdas} there are three different
regions corresponding to the different physical situations.
The most interesting
region is the ``Different" region where entanglement properties
of the inverted state are different from that of the initial one.
This means that {\it any} separability criterion which uses only
the symmetric CM will not detect these states, as the symmetric
CM is compatible with a separable as well as with an entangled state.
These states will not be detected by the CMC, and also not
by a variety of other criteria, as we will see later.


\subsection{Properties of covariance matrices}

In this section we will prove several properties of CMs which
are important for our later discussion. This concerns mainly
properties of CMs for pure states and the behavior of CMs under
the mixing of states.
We will first show in the subsequent
proposition that a suitable choice
of observables can dramatically simplify the form of CM $\gamma$
for pure states.

\begin{proposition}[Covariance matrices of pure states]
\label{CMform}
Let $G_i$ be a tomographically complete set of observables of a
$d$-dimensional system. If $\rho$ is a pure state then
$\gamma $ (as a $d^2\times d^2$ matrix) fulfills:
\bi
\item[{(i)}] The rank is given by $Rank(\gamma )=d-1$.
\item[{(ii)}] The nonzero eigenvalues of $\gamma$ are equal to 1,
hence $\tr(\gamma )=d-1.$
\item[{(iii)}] Consequently, we have $\gamma^2=\gamma$.
\ei
\end{proposition}
{\it Proof:} Without any loss of generality we assume $\rho=\ketbra{1}$
and take as observables the ones of the standard basis (\ref{standbas}).
Calculating directly and reordering of the matrix elements afterwards
gives a block structure \cite{blockremark}
\be
\gamma
=\bigoplus_{k=1}^{d-1}\big[B_k\big] \bigoplus \mathbbm{O}_{d^2-2d+2}
\;\mbox{ with }\;
B_k =
\begin{pmatrix}
1/2 & i/2
\\
-i/2 & 1/2
\end{pmatrix},
\ee
where $ \mathbbm{O}_{k}$ denotes a $k\times k$ matrix of zeros.
This matrix has the desired properties.
$\qed$

{From} this we can directly read off the properties of the
symmetric form of the covariance matrix:

\begin{corol}[Properties of symmetric CMs for pure states]
\label{CMformsym}
Let $\{G_i\}$ be a tomographic complete set of observables of a
$d$-dimensional quantum system. If $\rho$ is a pure state, then
$\gamma^S$ (as $d^2\times d^2$ symmetric matrix) fulfills:
\bi
\item[{(i)}] The rank is given by $Rank(\gamma^S)=2(d-1)$.
\item[{(ii)}] The nonzero eigenvalues of $\gamma^S$ are equal to $1/2,$
hence
$\tr(\gamma)=d-1.$
\ei
\end{corol}

We now turn to a proposition concerning the trace of a CM
for mixed states.

\begin{proposition}[Trace of CMs]
\label{trgthreshold} Let $\rho$ be a mixed state. Then
\be
\tr[\gamma({\rho})]= d-\tr(\rho^2)
\ee
which implies that
$d-1/d \geq \tr(\gamma({\rho}))
\geq d-1$. This holds also for $\gamma^S$.
\end{proposition}
{\it Proof:} By definition
$
\tr(\gamma)= \sum_i \gamma_{i,i}= \sum_i \delta^2(M_i)=
\sum_i(\mean{M_i^2}-\mean{M_i}^2).
$
The first summation is trivial, since we have
$\sum_k M_k^2=d\eins$ \cite{oPRA}. Furthermore
we can write $\rho=\sum_k \mean{M_k}M_k$ which implies that
$
\sum_k \mean{M_k}^2 =\tr(\rho^2),
$
and further $1/d \leq \tr(\rho^2) \leq 1.$
The statement for $\gamma^S$ follows directly
from the fact that $\tr(\gamma)=\tr(\gamma^S).$
$\qed$

We can also estimate the operator norm (i.e., the maximal
eigenvalue) of CMs.

\begin{proposition}[Operator norm of CMs]
For the CM $\gamma(\rho)$ and its linear part $\fg(\rho)$
the operator norm is bounded by
\be
\|\fg(\rho)\|\leq \|\rho\| \mbox{ and } \|\gamma(\rho)\|\leq \|\rho\|.
\ee
The same bounds hold for symmetric CMs.
\end{proposition}
{\it Proof:} Let us first consider $\fg(\rho).$ We have
$\|\fg(\rho)\| = \max_{\ket{x}} \bra{x}\fg(\rho)\ket{x}
= \bra{x_0}\fg(\rho)\ket{x_0}
= \tr(\rho A A^\dagger)$ with $\tr(AA^\dagger)=1.$
This is clearly smaller than $\|\rho\|.$ For $\gamma(\vr)$
this follows then from
$\mean{AA^\dagger}-\mean{A}\mean{A^\dagger}\leq \mean{AA^\dagger}.$
$\qed$

Finally, CMs also satisfy an interesting majorization relation.
This has its root in the way how one can relate CMs
to the rotated CMs of the pure states occurring in their
convex decompositions in terms of pure states.

\begin{proposition}[Majorization relation for CMs]
\label{majorization}For any (mixed) state $\rho$,
both the linear part $\fg(\rho)$ with entries
$\fg_{i,j}=\mean{M_iM_j}$ as well as the
CM $\gamma(\rho)$ satisfy
    \begin{equation}
        \sum_{j=1}^k \lambda_j[\fg(\rho)] , \sum_{j=1}^k
         \lambda_j[\gamma(\rho)]\leq \min(k,d-\delta_{\gamma}\frac{1}{d}),
    \end{equation}
    for the non-increasingly ordered eigenvalues, where
    $\delta_\gamma=1$ for $\gamma(\rho)$ $\delta_\gamma=0$
    if $\fg(\rho)$ is considered.
\end{proposition}
{\it Proof:} This is a consequence of
$\|\gamma(\rho)\|, \|  \fg(\rho) \|\leq 1$ as well as
of $\tr[\gamma(\rho)]\leq d-\frac{1}{d}$ and $\tr(\fg(\rho))\leq d$.
$\qed$

\subsection{Mixing property of covariance matrices}
Separable states are those states that can be written as convex
combinations of product states.
Therefore we have to understand the behavior of CMs under
mixing of states for the derivation of separability criteria.
An important property of covariance matrices which we refer to as
{\it concavity property} is the following:

\begin{proposition}[Concavity property]
\label{gammaprop}
Let $\rho=\sum_k p_k \rho_k$ be a convex combination of
states $\rho_k$, then
\be
\gamma(\rho) \geq \sum_k p_k\gamma(\rho_k).
\ee
Clearly, this implies the same relation for the symmetrized
CM $\gamma^{S}.$
\end{proposition}
{\it Proof:} As shown in Ref.\ \cite{oPRL} this inequality holds for
an arbitrary symmetric CM $\gamma^{S}$. Moreover, since
$
\mean{M_iM_j}_{\rho}
= \sum_kp_k\mean{M_iM_j}_{\rho_k}
$
for all $i,j$, we have for the non-linear part that
\begin{equation}
-\mean{M_i}_{\rho} \mean{M_j}_{\rho} \geq
-\sum_k p_k\mean{M_i}_{\rho_k}\mean{M_j}_{\rho_k}
\end{equation}
as a matrix inequality for the matrices $X_{i,j}=-\mean{M_i}_{\rho} \mean
{M_j}_{\rho}$
and $Y_{i,j}=-\sum_k p_k\mean{M_i}_{\rho_k}\mean{M_j}_{\rho_k}$.
{From} this, the above
inequality follows for the non-symmetric
CM $\gamma$.
$\qed$

This property will later be used to derive the separability criterion.

\subsection{Transformations of observables and validity of covariance matrices}

Transformations as generated by a general orthogonal
matrix $O$ used in Proposition
(\ref{bastrans}) do in general not preserve the positivity of the
state $\rho$ (see Ref.\ \cite{YuLiu} for discussion). Only a subgroup
will correspond to unitary transformations on the level of states.
Here, we will clarify how unitary transformations of the state
are reflected by orthogonal transformations on
the level of CMs.

For this aim, let us consider the case that $\rho$ is transformed
by some unitary transformation $\rho \mapsto U^\dagger \rho U$.
Equivalently, we can transform the operator basis, denoted as
$\{G_i\}$, as
\be
G_i \mapsto H_i= UG_{i}U^{\dagger}=\sum_j O_{i,j}G_j.
\label{generatedO}
\ee
It is then
easy to see that the transformation of the CM is
\be
\gamma(\rho)\mapsto O\gamma(\rho)O^T=\gamma(U^{\dagger}\rho U),
\ee
We can now ask in what way
$O$ depends on $U$, and which orthogonal
$O\in O(d^2)$ correspond to a unitary $U\in U(d)$ acting
in state space as  described above. That is, we look for the group
representation of $U(d)$ in the space of
CMs (compare
also the metaplectic representation of
symplectic transformations in discrete Weyl systems,
see Ref.\ \cite{Vourdas}).
The following
theorem gives an answer to this question.

\begin{proposition}[Transformation laws for CMs]
\label{OUconnection}
Let $U\in U(d)$. Then the $O\in O(d^2)$
representing $U$ as described above
(\ref{generatedO}) is given by
\be
O = \Gamma^T(U^T\otimes U^{\dagger})\Gamma^{\ast},
\ee
where $\Gamma$ is a $d^2\times d^2$ square matrix constructed as
$\Gamma_{\alpha,\beta | i}=
G^{\alpha,\beta}_i=(\ket{G_1},\ket{G_2},\dots)$,
where we understand $\alpha ,\beta$ as a row index, $G^{\alpha,\beta}_i$ as
vectors and construct $\Gamma_{\alpha,\beta | i}$ from
them.
\end{proposition}
{\it Proof:} The proof is given in the Appendix.
$\qed$

It is an interesting open question to see how CMs are
transformed under general completely positive maps,
$\rho\mapsto \sum_i A_i\rho A_i^{\dagger}$,
where $\{A_i\}$ are Kraus operators,
directly expressed in terms of the Kraus operators.

At the very beginning we have given
two definitions of covariance matrices for the symmetric and
non-symmetric case (\ref{cmdef},\ref{cmdefsym}). We will
discuss this difference also later in the paper.
However at this stage we mention a single
connection between these two definitions for block CMs:

\begin{proposition}[Block forms of CMs under local basis transformations]
\label{symm-nonsymm}
It is not possible to achieve for the block CMs
$\gamma=\gamma^{S}$ via local basis transformations of the operator
basis.
The only states for which this relation holds have the
reduced states
$\rho_{A}=\tr_B(\rho)=\eins/d_A$ and $\rho_B=\tr_A(\rho)=
\eins/d_B$,
where $d_{A,B}$ are the dimensions of $\rho_{A,B}$. It follows that
$\gamma=\gamma^{S}$ cannot be achieved by local unitary operations
either.
\end{proposition}
{\it Proof:} The proof is given in the Appendix.
$\qed$

\section{The covariance matrix criterion for separability}
\label{sepcrit}
In this section, we introduce the covariance matrix criterion (CMC)
for separability as our main topic of this paper. This is
derived from the concavity property for CMs and the fact that
the block CM for product states is block diagonal.

\begin{proposition}[Covariance matrix criterion]
\label{CMcrit}
Let $\rho$ be a separable state and $A_i$ ($B_i$) be
orthogonal observables on $\HH_A$ ($\HH_B$), where
the dimensions of the Hilbert spaces are $d_A$ ($d_B$, respectively).
Define $M_i=A_i\otimes\eins $ for $i=1,\dots ,d_A^2$ and
$M_i=\eins\otimes B_i $ for $ i=d_A^2+1,\dots ,d_B^2+d_A^2.$
Then there exist pure states $\ket{\psi_k}\bra{\psi_k}$ for $A$
and  $\ket{\phi_k}\bra{\phi_k}$ for $B$ and convex weights $p_k$
such that if we define
$\kappa_A=\sum_k p_k \gamma(\ket{\psi_k}\bra{\psi_k})$
and
$\kappa_B= \sum_k p_k\gamma(\ket{\phi_k}\bra{\phi_k})$
the inequality
\be
\gamma^S(\rho,\{M_i\}) \geq\kappa_A\oplus\kappa_B
\Leftrightarrow
\begin{pmatrix} A & C\\
                C^T & B
\end{pmatrix}\geq
\begin{pmatrix} \kappa_A & 0 \\
                0        & \kappa_B
\end{pmatrix}
\label{CMC}
\ee
holds. This means that the difference between left and right hand side
must be positive-semidefinite. If there are no such $\kappa_{A,B}$ then
the state $\rho$ must be entangled.
\end{proposition}

{\it Proof:} First note that for this special choice of $M_i$, for any
product state
\be
    \gamma(\rho_A \otimes \rho_B, \{M_i\})=\gamma(\rho_A,\{A_i\})\oplus
    \gamma(\rho_B,\{B_i\})
\ee
holds. Now, since any separable state can be written as
$\rho = \sum_k p_k \ketbra{\psi_k} \otimes \ketbra{\phi_k},$
we can apply Prop.~(\ref{gammaprop}) and arrive at the
conclusion.
$\qed$

Note that the CMC is manifestly invariant under a change of the
observables $\{A_k\}$ and $\{B_k\}$,
as we know from Proposition \ref{bastrans}
[see also Eq.~(\ref{bastranseq})]; however,
a suitable choice of them may simplify the evaluation a lot. Also, note
that we have formulated the CMC for symmetric CMs, we will later
discuss the case of non-symmetric CMs.

Obviously, as such, as formulated as in Prop.~\ref{CMcrit}, it is
not clear that the CMC leads to an efficient and physically plausible test for
separability: The main problem is to characterize the possible $\kappa_A$
and $\kappa_B$. As such, the formulation still contains an optimization over
all pure product states. We will refer to an ``evaluation of the CMC'' hence
whenever we can identify a property of $\kappa_{A,B}$ that will render
the above criterion an efficient test.

Some properties of we have derived above, notably
\begin{equation}
    \tr(\kappa_A) = d_A -1
\end{equation}
(see Proposition \ref{CMformsym}),
which we will use subsequently.
We will now turn to feasible ways to evaluate the CMC. As a
first step, we have to derive conditions on the blocks of a
block matrix as in Eq.~(\ref{CMC}), which follow from the positivity
condition in Eq.~(\ref{CMC}). Then, we ask how the observables
$\{A_k\}$ and $\{B_k\}$ must be chosen in order to make a violation
of Eq.~(\ref{CMC}) manifest.

Note the formal similarity of
the condition $\gamma  \geq\kappa_A\oplus\kappa_B$
to tests for separability for {\it Gaussian
states} for systems with {\it canonical coordinates}. A CM
in that context \cite{Survey,Survey2}
is any symmetric matrix satisfying
$\gamma + i\sigma\geq 0$, where
\begin{equation}
    \sigma = \bigoplus_{k=1}^n \left(\begin{array}{cc}
    0 & 1\\
    -1 & 0\\
    \end{array}
    \right).
\end{equation}
A Gaussian state $\rho$ with CM $\gamma$
is now separable if and only if \cite{CVbe}
there exist covariance matrices $\gamma_A,\gamma_B$,
each satisfying $\gamma_A,\gamma_B+ i\sigma$
such that
\begin{equation}\label{sd}
    \gamma\geq \gamma_A\oplus\gamma_B.
\end{equation}
For non-Gaussian states, a violation of this condition is still
sufficient to detect entanglement. Eq.\ (\ref{sd}),
in turn, is simply a semi-definite program (SDP), so it can be
efficiently decided \cite{SDP}. The characterization of the right hand
side is here an easy task, as these matrices are again
constrained by a semi-definite constraint.
We will see in section \ref{sec:CMC42q} that for two qubits
the CMC can be solved in a similar way as in
the continuous variable case \cite{sdprog}.

\section{Evaluation of the CMC}

We would like to follow two strategies to evaluate the CMC
as presented in the previous section. Both
strategies are based on matrix invariants such
as eigenvalues or singular values. Let us first
characterize positive semi-definiteness of a block matrix
of the type as in (\ref{CMC}) in terms of singular
values of its submatrices.

\subsection{Evaluation of the CMC via singular values of submatrices}

As a start, we state the following lemma:

\begin{lem}[Block covariance matrices and unitarily invariant norms]
\label{BCMUIN}
If a positive matrix partitioned in block form is positive
semi-definite,
\be
\begin{pmatrix}A&C\\ C^T&B\end{pmatrix}\geq 0,
\label{claim}
\ee
then
\be
\NORM A \NORM\, \NORM B \NORM  \geq \NORM \, |C|\, \NORM^2,
\label{spurnormeq}
\ee
holds, where $\NORM . \NORM $ is any unitarily invariant
norm. Specifically, this holds true for any
Ky-Fan norm $\Vert .\Vert_{KF}$ defined as
the sum of the largest $k$ singular values. If we sum over
all singular values, we arrive at the largest Ky-Fan
norm, which is the trace norm.
\end{lem}

{\it Proof:} The proof of this statement is actually a
corollary of Theorem 3.5.15 of Ref.\ \cite{HornJohnson}.
It is shown that
\be
\NORM A^p \NORM\, \NORM B^p\NORM  \geq \NORM(C^\dagger C)^{p/2} \NORM^2,
\ee
for any $p>0$ and any unitarily invariant norm. For $p=1$
this is the result we are interested in. We will nevertheless
present an alternative proof of this statement for Ky-Fan norms
$
\Vert A \Vert_{KF} \Vert B\Vert_{KF}  \geq \Vert C \Vert_{KF}^2,
\label{spurnorm}
$
in the Appendix, as the proof of Proposition \ref{coptimal}
will make use of this proof.\qed

Using the last Lemma we have:

\begin{proposition}[CMC evaluated using singular values]
\label{zhangcrit}
Let
\be
\gamma=\begin{pmatrix} A& C\\ C^T & B \end{pmatrix}
\ee
be a CM. Then, if $\rho$ is separable,
we have
\be
\Vert C \Vert_{Tr}^2 \leq {\left[1-\tr\left(\rho_A^2\right)\right]
                     \left[1-\tr\left(\rho_B^2\right)\right]}.
\ee
If this inequality is violated, then $\rho$ must be entangled.
\end{proposition}

{\it Proof:} We prove the claim applying the formula (\ref{spurnormeq})
directly, yielding
\begin{equation}
\Vert C\Vert^2_{Tr}\leq \Vert A-\kappa_A \Vert_{Tr}\Vert B-\kappa_B
\Vert_{Tr}
\end{equation}
Since $A-\kappa_A$ as well as $B-\kappa_B$ are Hermitian
positive semi-definite matrices (due to concavity property
of CMs) their trace norm will coincide with their trace.
Hence
$
\Vert A-\kappa_A \Vert_{Tr}=\tr(A)-\tr(\kappa_A)=1-\tr(\rho^2_A),
$
where we have used Corollary \ref{trgthreshold} and the fact
that $\sum_{i}A_{i,i} = \sum_{i} (\mean{A^2_i}-\mean{A_i}^2) = \mean{d_
A\Eins} - \tr(\rho^2_A)$,
since $\tr(A_iA_j)=\delta_{i,j}$ and $\rho_A^2=\sum_{i,j} \mean{A_i}
\mean{A_j}A_iA_j$.
$\qed$

Interestingly, this criterion has been proven
already in a different context:

\begin{rem}[CMC and the criterion of Ref.\ \cite{china2}]
The separability criterion in Proposition \ref{zhangcrit}
is nothing but the separability criterion proposed in Theorem 1
of Ref.~\cite{china2}, hence the criterion of  Ref.~\cite{china2}
is a corollary
of the CMC.
\end{rem}

Let us now connect the CMC to another type of entanglement
criteria: There are several separability criteria in the
literature which are based on the Bloch representation of
density matrices. This representation in our case is just
some particular choice of observables, namely one has to
detach the identity from all others generators, which then
have to be traceless. The fact that one of the observables
is the identity, can simplify the CMC sometimes.

By definition the entries of the matrix $C$
are given by
\be
C_{i,j}=\mean{A_i\otimes B_j} - \mean{A_i}\mean{B_j},
\ee
which consists of a
linear (in the sense of mean values) and quadratic part. We
define $\fC$ as the linear part of $C$, i.e., $\fC_{i,j}=\mean{A_i\otimes
 B_j}$.
Let us further consider $\fC^{\text{red}}$ as the submatrix of $\fC$, where
the entries $\mean{\Eins_A \otimes B_j}$ and
$\mean{A_i\otimes\Eins_B}$ are omitted,
i.e., the first row and the first column are removed.
Similarly, we can define matrices like
$\fA$ and  $\fB^{\text{red}}$ from $A$ and $B.$ In the same spirit, we can
define a submatrix of $\kappa$ as $\kappa^{\text{red}}.$ Note that
$\tr(\kappa) = \tr(\kappa^{\text{red}}),$
as the missing diagonal entry is the variance of $\openone,$ which is
vanishing.

We are now able to establish a connection between the CMC and
criteria based on the Bloch representation of density matrices:

\begin{proposition}[Relationship between CMC and criteria based on Bloch
representations]
\label{vicentecrit}
Let
\be
\gamma=\begin{pmatrix} A& C\\ C^T & B \end{pmatrix}
\ee
be a CM. Then if $\rho$ is separable, we have
\be
\Vert \fC^{red}\Vert_{Tr}^2
\leq
{\big(1-\frac{1}{d_A}\big)
\big(1-\frac{1}{d_B}\big)}.
\ee
If this inequality is violated, then $\rho$ must be entangled.
\end{proposition}

{\it Proof:} First, we can define two vectors
$\ket{\psi}^{A/B}$ with entries
\be
\ket{\psi^A}_i = \mean{A_i}
\;\;\;
\ket{\psi^B}_i = \mean{B_i}
\ee
resulting in  $C = \fC - \ket{\psi^A}\bra{\psi^B}$.
Similar relations hold for $A$ and $B$, so we can write
the condition in CMC
(\ref{CMC}) in the form
\be
\underbrace{\begin{pmatrix} \fA - \kappa_A & \fC\\
                \fC^T & \fB - \kappa_B
\end{pmatrix}}_{X} -
\begin{pmatrix}\ket{\psi^A}\\ \ket{\psi^B}\end{pmatrix}
\begin{pmatrix}\bra{\psi^A}\\ \bra{\psi^B}\end{pmatrix}^T
\geq 0.
\ee
Positivity of the left hand side implies positivity of the first term $X$
alone, since we subtract only one projector which is itself positive.
Concerning positivity of $X$ we can take $\fA^{\text{red}}$,
$\fB^{\text{red}}$, $\fC^{\text{red}}$,
and $\kappa^{\text{red}}$ instead, since positivity
of
a matrix implies positivity of all its main minors.
Using Eq.~(\ref{spurnormeq}), we get
\be
\Vert \fC^{\text{red}}\Vert_{Tr}
\leq
\Vert \fA^{\text{red}}-\kappa_A^{\text{red}}\Vert_{Tr}
\Vert \fB^{\text{red}}-\kappa_B^{\text{red}}\Vert_{Tr}.
\ee
Using that
$\tr(\hat\fA^{\text{red}}) =
\sum_{i\geq 2} \mean{A^2_i} =
\mean{d_A \Eins_A} - \mean{\Eins_A / d_A}$ and
$\tr(\kappa_A^{\text{red}})=d_A-1$ proves the claim.
$\qed$

Interestingly, this separability criterion has also been
proven before:

\begin{rem}[CMC and the criterion of Ref.\ \cite{deVic}]
The separability criterion in Proposition \ref{vicentecrit}
is nothing but the separability criterion for Bloch representations
proposed in  Ref.~\cite{deVic}, hence the criterion of
Ref.~\cite{deVic} is a corollary of the CMC.
\end{rem}

Note that in Ref.~\cite{deVic} the observables have been normalized in
a different way, leading to a slightly different formula.

\begin{rem}[Connection between Propositions \ref{vicentecrit} and
\ref{zhangcrit}]
Proposition \ref{zhangcrit} is strictly stronger than Proposition \ref{vicentecrit}.

This fact was proven in version 5 of ~\cite{china2}, which came out after
submission of our paper to the arXiv.
\end{rem}

Let us finish this subsection with a remark on
the possible use of other Ky-Fan norms in the above
argument. In fact, we do know  more about the singular
values (here eigenvalues)
of $\kappa_A$ and $\kappa_B$ than their sum:

\begin{lem}[Ky-Fan norms of matrices in the CMC]
The matrices $\kappa_A$ (and similarly
$\kappa_B$) in Proposition \ref{CMcrit} satisfy
\begin{equation}
    \sum_{j=1}^k
    \lambda_j(\kappa_A)\leq \min(k,d_A-1),
\end{equation}
for the non-increasingly ordered eigenvalues of
$\kappa_A$ (and $\kappa_B$).
\end{lem}

{\it Proof:} One can argue as in Proposition
\ref{majorization}, using the fact that a convex combination
of matrices leads to more mixed matrices in the sense
of majorization \cite{HornJohnson}.
\qed

This property can immediately be applied to
evaluate the CMC, making use of proposition
\ref{BCMUIN} and Weyl's inequalities \cite{WE}.
For example, if we consider $d_A=d_B=d$
and the $k$-Ky-Fan norm $\|.\|_{KF^{(k)}}$ for $k=(d^2-d+1+s)$,
we can apply the first of Weyl's inequalities with $i=1$,
and $s=1,\dots, d-1$,
to conclude that
\begin{eqnarray}
    \|A-\kappa_A\|_{KF^{(k)}} &=&
    \sum_{j=1}^k
    \lambda_j (A-\kappa_A)\nonumber\\
    &\leq &
    \sum_{j=1}^k
    \lambda_1 (A)+ \sum_{j=1}^k
    \lambda_j (-\kappa_A)\nonumber\\
    &=&(d^2-d+1+s)
    \|A\|\nonumber\\
    &-& \sum_{j=d^2-k+1}^{d^2}
    \lambda_j (\kappa_A),
\end{eqnarray}
where $\|A\|$ denotes the spectral norm of $A$.
Using that $\kappa_A$ will be more mixed
in the sense of majorization than $diag(1,\dots, 1,0,\dots, 0)$
of rank $d-1$ and Proposition \ref{trgthreshold}, one arrives at
\begin{eqnarray}
    \|A-\kappa_A\|_{KF^{(k)}} &\leq&
    (d^2-d+1+s)
    \|A\|-s,
\end{eqnarray}
and a corresponding statement for $\kappa_B$.
Using Proposition \ref{BCMUIN}, one hence arrives
at the observation that
any separable state $\rho$ on a bipartite
Hilbert space satisfies
\begin{eqnarray}
    &&\left((d^2-d+1+s)\|A\|-s\right)
    \left((d^2-d+1+s)\|B\|-s\right)\nonumber\\
    &-& \|C\|_{KF^{(k)}}^2\geq 0.
\end{eqnarray}
It is an interesting open question whether more
sophisticated uses of the knowledge of spectral properties
of $\kappa_A$ and $\kappa_B$ can be employed
the further sharpen the evaluation of the CMC.

\subsection{Evaluation of the CMC via traces of
submatrices}

Let us first prove a simple condition on the traces of
$A,B$ and $C,$ which follows from the CMC. In the
following, we always assume that $d_A \leq d_B.$ Sometimes
we assume that the dimensions are the same, meaning that
$d=d_A=d_B.$

\begin{proposition}[CMC evaluated from traces]
\label{generalcrit}
Let
\be
\gamma=
\begin{pmatrix}
A& C\\
C^T & B
\end{pmatrix}
\ee
be the symmetric CM  of a state $\rho$ and let
$J=\{j_1,\dots,j_{d_A^2}\}\subset\{1,\dots,d_B^2\}$
be a subset of $d_A^2$ pairwise different indices.
Then if $\rho$ is separable, we have
\begin{align}
2 \cdot  & \sum_{i=1}^{d_A^2} \sum_{j\in J}|C_{i,j}| \leq
\Big(\sum_{i=1}^{d_A^2}A_{i,i}-d_A+1\Big)+
\\
&+\Big(\sum_{i=1}^{d_B^2}B_{i,i}-d_B+1\Big)
=\big[1-\tr(\rho_A^2)\big]
+\big[1-\tr(\rho_B^2)\big],
\nonumber
\end{align}
If this inequality is violated, then $\rho$ must be entangled.
\end{proposition}

{\it Proof:} First, note that a necessary condition for a
$2\times 2$ matrix
\be
X=
\begin{pmatrix} a & c\\ c &b\end{pmatrix}
\ee
to be positive semidefinite is that $2|c|\leq a+b$.
If $\rho$ is separable, then by the CMC we have
$Y=\gamma-\kappa_a\oplus\kappa_B\geq 0$. This implies
that all $2\times 2$ minor submatrices of $Y$ have to
be positive semidefinite as well. Hence for all
$i,j$ we have
\be
2|C_{i,j}|\leq A_{i,i}+B_{j,j} - (\kappa_A)_{i,i} - (\kappa_B)_{j,j}.
\label{tosum}
\ee
Summing over $i,j$ and using Corollary \ref{trgthreshold} proves
the claim.
$\qed$

We will use this Proposition mainly for the case that $d_A= d_B$
and where we sum over the diagonal entries of $C.$ In this case,
it just gives the condition that for separable states,
$2 \tr(C) \leq 2 - \tr(\rho_A^2) - \tr(\rho_B^2)$. This is a
quadratic polynomial in the entries of the state, and may be viewed
as a suitable entanglement witness on two specimens on a
state. In the light of this fact, the criterion evaluated in this
fashion is surprisingly strong.
It is also worth mentioning here that one can improve
Proposition \ref{generalcrit} by taking $4\times 4$ minor
 submatrices for evaluation. Then, however, also off diagonal
terms of $\kappa_{A/B}$ will occur, for which not many
properties are known. This makes the resulting conditions
difficult to evaluate.

Physically, Proposition \ref{generalcrit} says that if the
correlations $C_{i,j}$ are sufficiently large,
then $\rho$ must be
entangled. The question arises, how to find the observables,
for which the $C_{i,j}$ are large. There are several ways of
doing this. A first result is the following:

\begin{proposition}[Criterion in Proposition
\ref{generalcrit} and diagonal block matrices]
\label{coptimal}
The criterion in Proposition
\ref{generalcrit} detects most states if the observables
are chosen in such a way that $C$ is diagonal. For any
state there exist a choice of observables that this can
be achieved.
However, even with this optimal choice of observables Proposition
\ref{generalcrit} delivers a strictly weaker separability criterion
than Proposition \ref{zhangcrit}.
\end{proposition}

{\it Proof:} It is clear that the criterion is
optimal, if the  trace of $C$ is maximal, which
is the case if it is brought into the singular value
diagonal form \cite{china2, wir}. This can always be achieved
[see Eq.~(\ref{bastranseq})]. The fact that Proposition
\ref{zhangcrit} is stronger, was in a different language proven in
Ref.~\cite{china1}.

Interestingly, the fact that Proposition \ref{generalcrit} is weaker than
Proposition \ref{zhangcrit} can also be seen from
Eq.~(\ref{interestingequation}) from the alternative proof of Lemma \ref{BCMUIN}.
If $C$ is chosen to be diagonal, then Proposition
\ref{generalcrit} reduces to this equation with $\alpha=\beta.$ Clearly,
allowing
$\alpha$ and $\beta$ to be different, improves the criterion.
$\qed$

In the following, however, we will consider two
different strategies: Firstly, we use the Schmidt
decomposition in operator space  of the density matrix
\cite{oPRA}. This will lead to a natural choice of the
observables $\{A_k\}$ and $\{B_k\}$, and will further connect
the CMC to the CCNR criterion.

Secondly, we will
consider appropriate {\it local filterings} of the
state \cite{norway, frankfilterqubit, frankfiltermulti, filtop}.
These are active transformations of the state,
which, however, do not change the entanglement properties.
Under this transformations,
the state can be transformed into what is called its
{\it standard form} in Ref.~\cite{norway}. In this standard
form, the CMC becomes very strong and even necessary and sufficient
for two qubits.

\subsection{Schmidt decomposition and the CMC}

We will first remind ourselves of what is called the {\it Schmidt
decomposition in operator space}. It is the same construction
as the ordinary Schmidt decomposition in the vector space now
equipped with the Hilbert-Schmidt scalar product.
A general density matrix of a composite system can be written as
\be
\rho = \sum_{k=1}^{d_A^2}\sum_{l=1}^{d_B^2} \xi_{k,l} \tilde{G}_k^A
\otimes \tilde{G}_l^B,
\label{rhoinbasis}
\ee
with real $\xi_{k,l}$ and the $\{\tilde{G}_l^A\}$ (respectively, $\{\tilde{G}_l^B\}$)
form an orthonormal Hermitian basis
of observables. The Schmidt decomposition can now be achieved by
diagonalizing the above expression using
the singular value decomposition of the matrix $\xi$,
\be
\rho = \sum_{k=1}^{d_A^2}\lambda_{k} G_k^A\otimes G_k^B,
\ee
where we made the assumption that $d_A\leq d_B$. Clearly, the Schmidt
coefficients
$\lambda_k$ are real and non-negative. Using the new basis observables
$\{G_k^A\}$ and $\{G_k^B\}$ as observables for the construction of
the symmetric block CM,
we have a normal form of the CMC, which we will call the
{\it Schmidt CMC}.

\begin{proposition}[Schmidt CMC]
\label{schmidtdec}
If $\rho$ is separable, then
\be
2\sum_i|\lambda_i-\lambda_i^2g_i^Ag_i^B|\leq
2-\sum_i\lambda_i^2\left[(g_i^A)^2+(g_i^B)^2\right],
\ee
where we defined $g_i^A = \tr(G_i^A)$ and
$g_i^B = \tr(G_i^B)$. If this condition is violated,
the state must be entangled.
\end{proposition}

{\it Proof:}
Using the orthonormality of the $\{G_i^{A/B}\}$,
it is not difficult to see that with the
observables from the Schmidt decomposition
$
C_{i,j}=\lambda_{i}\delta_{i,j}-\lambda_{i}\lambda_{j} g_j^A g_i^B$
holds. In addition, we have
$\tr(\rho_A^2)=
\sum_i
\lambda^2_{i} (g_i^B)^2.
$
Together with Proposition \ref{generalcrit} this proves the claim.
$\qed$

Interestingly, this Proposition includes the CCNR criterion as a
corollary. This shows that the CMC, even without filtering, and
evaluated merely via the trace of the blocks, once the matrix
is brought to Schmidt form, is stronger than
the CCNR, which it implies as a corollary.

\begin{corol}[CMC and CCNR]
If a state $\rho$ is separable, then in the Schmidt decomposition
\be
\sum_k \lambda_k \leq 1
\ee
has to hold. This condition is just the CCNR criterion, hence the
CCNR criterion is a corollary of the CMC.
\end{corol}

{\it Proof:}
Using  the relations $|a-b|\geq |a|-|b|$ and $a^2+b^2\geq 2|ab|$
we have
$
2\sum_i|\lambda_{i}-\lambda^2_{i}g_i^Ag_i^B|
\geq
2\sum_i\lambda_{i}-2\sum_i\lambda^2_{i}|g_i^Ag_i^B|
$
and
$
2 - \sum_i\lambda^2_{i}[(g_i^A)^2+(g_i^B)^2]
\leq
2(1-\sum_i\lambda^2_{i}|g_i^Ag_i^B|)$,
which, due to the Proposition \ref{schmidtdec},
proves the claim.
$\qed$

\subsection{Filtering and the CMC}
Let us now consider local filtering operations or SLOCC
(stochastic local operations assisted by classical
communication) \cite{filtop} of the form
\be
    \rho\mapsto\tilde{\rho}=(F_A\otimes F_B)
    \rho (F_A\otimes F_B)^{\dagger},
\ee
where and $F_A\in\mbox{SL}(d_A,\C)$ and $F_B\in\mbox{SL}(d_B,\C)$ are
invertible matrices on the respective Hilbert spaces. Clearly, such
operations cannot map a separable state into an entangled one
(although they might increase entanglement measures).
Also, since $F_A$ and $F_B$ are
invertible, they will also not destroy any entanglement that may be
present in the state. In other words, $\rho$ is entangled
if and only if $\tilde{\rho}$ is entangled.

As has been shown in Refs.~\cite{filtop, norway} we can
bring any state of full rank (i.e., $\rho > 0$)
by such filtering operations in its {\it standard form} which
is given by
\be
\label{standform}
\tilde{\rho}=
\frac{1}{d_A d_B}
\Big(
\eins+\sum_{k=1}^{d_A^2-1}\xi_k\gh_k^A\otimes\gh_k^B
\Big)
\ee
where the $\{\gh_k^A\}$, $\{\gh_k^B\}$ are {\it traceless} orthogonal
observables. Here, we again assumed that $d_A\leq d_B$.

The idea now is to first apply a filtering operation and bring the state
into its normal form. Then, the new separability criteria
are applied afterwards. Note that the reduction to the normal
form is always possible.
The merits of this approach are twofold:
Firstly, the normal form
reduces the number of relevant parameters, while still encoding all
information about entanglement and separability. Secondly,
the normal form is in a certain sense ``more entangled'' than
the original state, as it was shown
in Ref.~\cite{frankfiltermulti}:

\begin{rem}[Extremality of states in normal form]
The local filtering operations bringing a mixed state into its normal
form are those operations
which maximize all entanglement monotones that
remain invariant under determinant 1 SLOCC operations.
\end{rem}

Therefore, it may be expected that many separability criteria
become stronger if we first bring the state into its normal form.
Note, however, that this does not hold for the PPT criterion, as
local filtering leaves this criterion invariant.

Following Ref.\ \cite{norway}, let us explain briefly an algorithm for
transforming a state of a form as in Eq.~(\ref{rhoinbasis}) to
its normal form  in Eq.~(\ref{standform}). As a starting
point, one considers the compact space $D_A\otimes D_B$ of all
normalized product density matrices $\rho_A\otimes\rho_B$.
For any given density matrix $\rho$ one can define a function
$f$ of $\rho_A$ and $\rho_B$ via
\be
f_{\rho}(\rho_A,\rho_B)
=\frac{\tr\left[\rho (\rho_A\otimes\rho_B)\right]}
{\left( \det\rho_A\right)^{1/d_A}
\left( \det\rho_B\right)^{1/d_B}
}.
\label{fdef}
\ee
$f_{\rho}(\rho_A,\rho_B)$ is a family of positive
well defined functions on the interior of
$D_A\otimes D_B$, where the reduced density matrices
both have full rank. Since $\rho$ has also full rank,
we have
$
\tr\left[\rho (\rho_A\otimes\rho_B)\right]>0
$
and because of compactness of $D_A\otimes D_B$ one has
even stronger
$
\tr\left[\rho (\rho_A\otimes\rho_B)\right]\geq c_{\rho}>0.
$
Divergence of $f_{\rho}(\rho_A,\rho_B)$ on the boundary
implies that it has a positive minimum
on the interior of $D_A\otimes D_B$.

Minimization of the function $f_{\rho}$ will, as proven
in Ref.\ \cite{norway}, yield the filtering operations needed.
Suppose the minimum value for $f_{\rho}$ attained for
some product density matrix $\tau_A \otimes \tau_B$
with $\det\tau_A>0$, $\det\tau_B>0$. Each of them can be
decomposed as (see Eq.~(66) in Ref.\ \cite{norway})
\be
\tau_A =
T_A^{\dagger}T_A,
\quad
\tau_B = T_B^{\dagger}T_B,
\quad T_{A/B}\in
\mbox{SL} (d_{A/B},\mathbb{C})
\ee
where the $T_A$ and $T_B$ are desired local filtering operations.
Normalization factors have been ignored.

Using this filtering operations one obtains the new state
$\tilde{\rho}$ which has a form
\be
\tilde{\rho}=\frac{1}{d_Ad_B}
\Big(
\eins+\sum_{i=1}^{d_A^2-1}
\sum_{k=1}^{d_B^2-1}\xi_{ik}\gh_i^A\otimes\gh_k^B
\Big)
\ee
The final step involves a standard singular value
decomposition of $\xi_{ik}$, which leads
to Eq.\ (\ref{standform}). A priori, it is not clear
whether the normal form is in some sense unique or not.
However, it is easy to see that if we start from a given state
and convert it into two different states in a normal form,
then these two normal forms have to be connected by a local
filtering operation. Using the fact that the reduced states
of a state in the normal form are maximally mixed,
one can further conclude that two different normal forms
can only differ by a local unitary transformation.

In practice, the minimization of $f_{\rho}(\rho_A,\rho_B)$
in Eq.~(\ref{fdef}) can be performed by an iteration as
follows: let us  fix $\rho_B$ and consider only
the minimization over $\rho_A.$ This minimization can
further be split into a minimization over the spectrum
of $\rho_A$ and a local unitary transformation.
If the spectrum is fixed, the optimal unitary is
constructed such that $\rho_A$ and $X=\tr_B[\rho
(\openone \otimes \rho_B)]$ are diagonal in the same
basis where the maximal eigenvalue of $X$ has the same
eigenvector as the minimal eigenvalue of $\rho_A$ and
the second largest eigenvalue of  $X$ has the same
eigenvector as the second smallest eigenvalue of $\rho_A$ etc.
If the basis is fixed, and $\lambda_k$
($\mu_k$) are the eigenvalues of $\rho_A$ ($X$) then
a simple calculation using Lagrange multipliers shows that
the optimal $\lambda_k$ fulfill
\begin{equation}
    \lambda_k \sim \biggl[{(\sum_{i \neq k}\mu_i \lambda_i)/
    (\prod_{i\neq k}\lambda_i)}\biggr]^{1/2},
\end{equation}
which can be used for an iterative determination of the
optimal $\lambda_k$. In this way, the optimization can be
iterated, converging to a minimum.
Note while it is known that every state can be
brought into this normal
form, the above procedure of Ref.\
\cite{norway} is not known to be strictly efficient in the
physical dimension $d$. Yet, for ``reasonable physical dimensions'', the method in practice converges quickly.
Moreover, and importantly,
at the end of the procedure, one can easily (and efficiently) check
via direct inspection whether the obtained filters map the state onto the
normal form or not.
Global optimality can hence be easily certified.


As one can directly calculate, for a state in the normal form the
CM takes a really simple form, namely
\be
\gamma =
\frac{1}{d_Ad_B}
\left(\begin{array}{cc}
\text{diag}(0,d_B,d_B,\dots ) & \text{diag}(0,\xi_1,\xi_2,\dots)\\
\text{diag}(0,\xi_1,\xi_2,\dots) & \text{diag}(0,d_A,d_A,\dots )
\end{array}\right).
\label{ccngamnonsym}
\ee
Using this form we obtain the following strong separability
criterion, which we call
the {\it filter CMC}.

\begin{proposition}[Filter CMC]\label{cmcnfgamsym}
If $d=d_A=d_B$ and $\rho$ is separable, then the
coefficients in
the filter normal
form fulfill
\be
\sum_i\xi_i\leq d^2 - d.
\ee
\end{proposition}

{\it Proof:} The claim obviously follows from Proposition
\ref{generalcrit} and the form of
the CM for the normal form of the state.
$\qed$

Interestingly, for two qubits we have:

\begin{rem}[Filter CMC for two qubits]
For two qubits, the filter CMC in Proposition \ref{cmcnfgamsym} is a
necessary and
sufficient criterion for separability.
\end{rem}

{\it Proof:}
If a two-qubit state is of full rank, the normal
form reads
\be
\tilde{\rho}=\frac{1}{4}\Big(\eins+\sum_{k=1}^{3}\xi_k\sigma_k^A\otimes
\sigma_k^B\Big),
\ee
where $\{\sigma_k^{A/B}\}$ are the
Pauli matrices \cite{norway}. Such states
are diagonal in the Bell basis, and it is known that for these states
$\sum_{k=1}^{3} \xi_k \leq 2$ is necessary and
sufficient for separability \cite{HH,norway}. Note also that the
filter normal form can be explicitly stated for two-qubit systems.

If an entangled (or separable) state is not of full rank, it can, as
explicitly shown  in Ref.~\cite{frankfilterqubit}, be brought by
filtering operations arbitrarily
close to a Bell diagonal state with finite (or vanishing) concurrence.
Such a state will also be detected by the CMC (or not).
$\qed$

%

Direct comparison of this result with the discussion in
Section II and Fig.~\ref{lambdas} (and later the result
of Proposition \ref{LURCMC}) might be confusing at this point,
since we know already that the CMC itself cannot be necessary
and sufficient for two qubits. This can be resolved in the following way:
We have already learned that filtering
brings the state in the form which in a certain sense contains
the maximum amount of entanglement (it maximizes all monotones).
This indeed shows that the filter CMC is sometimes
a real improvement of the ``bare'' CMC, and filtering is more
than just an appropriate choice of the observables.

Let us now consider the asymmetric case, when $d_A < d_B$.
We can formulate for this case following statement:

\begin{proposition}[Separability criterion for uneven local dimension]
\label{uneven}
If $\rho$ is separable, then the following inequalities hold
\begin{align}
\sum_i\xi_i&\leq \frac{d_Ad_B}{2}\Big[ 1-\frac{1}{d_A}+(d_A^2-1)\frac{1}{d_
B}
\nonumber
\\
&+\min (0,
-(d_B-1)+(d^2_B-d^2_A)\frac{1}{d_B})\Big]
\label{nfcritnonsym}
\end{align}
and
\be
\sum_i\xi_i\leq \left[{d_Ad_B(d_A-1)(d_B-1)}\right]^{1/2}
\label{eq3}.
\ee
holds. If one of these inequalities is violated,
the state must be entangled.
\end{proposition}

{\it Proof:} Eq.~(\ref{eq3}) is nothing but an application of
Proposition \ref{vicentecrit}  (or \ref{zhangcrit}),
it has already
been derived in Ref.\ \cite{deVic}.
Concerning Eq.~(\ref{nfcritnonsym}), we will again apply Proposition
\ref{generalcrit}, but with two modifications. First, when carrying out
the sum  over $2|C_{i,j}|\leq A_{i,i}+B_{j,j} - \kappa_{A,i,i} -
\kappa_{B,j,j}$
[see Eq.~(\ref{tosum}) in Proposition \ref{generalcrit}] we
do not sum over all $B_{i,i}$.
But then, we cannot subtract all of the $\kappa_{B,i,i}$
anymore, since $d_B^2-d_A^2$ diagonal elements of $\kappa_B$ do
not occur in the sum.

As a first approach, we can drop completely the summation
over all $\kappa_{B,j,j},$ since they are positive anyway.
This gives
\be
\frac{2}{d_Ad_B}\sum_{i=1}^{d_A^2}\xi_i \leq
1-\frac{1}{d_A}+\frac{d_A^2-1}{d_B},
\label{kappazeroestim}
\ee
justifying one part of  Eq.~(\ref{nfcritnonsym}).

In a second approach, we estimate $\sum_{i=1}^{d_A^2}\kappa_{B,i,i}$.
As one can see by direct inspection,
the non-vanishing elements of $\gamma$ in
Eq.~(\ref{ccngamnonsym}) origin only from the linear part of CM
(in the spirit
of Proposition \ref{vicentecrit} this linear part is denoted by $\fg$).
But as we have seen  in the proof of Proposition \ref{gammaprop}
that
this linear part $\fg$ is just the same as the linear part of the
direct sum of $\kappa_A\oplus\kappa_B$ (denoted by $\fk_A \oplus \fk_B$)
for separable states, i.e.
$\fg = \fk_A \oplus \fk_B$,
hence
$B = \fB =  \fk_B$.
This implies that for the diagonal elements of $\kappa_B$ the relation
$\kappa_{B,i,i}
\leq
\fB_{i,i}=B_{i,i}={1}/{d_B}$
holds, leading to
\begin{eqnarray}
    \sum_{i=1}^{d_A^2}\kappa_{B,i,i}&=&
    d_B-1-\sum_{i=d_A^2+1}^{d_B^2}\kappa_{B,i,i}\nonumber\\
    &\geq&
    d_B-1-\left(d_B^2-d_A^2\right)\frac{1}{d_B}.
\end{eqnarray}
This proves the second part of Eq.~(\ref{nfcritnonsym}).
$\qed$

\section{Connection to local uncertainty relations}

In this section we will further
analyze the connection of CMC
with the separability criterion based on
{\it local uncertainty relations} (LURs) \cite{LUR}.
To start with, we again state the LUR criterion as a reminder:

\begin{proposition}[Criterion based on local uncertainty relations]
\label{LURs}
Let be $\hat{A}_k$ and $\hat{B}_k$ observables in system $A$ and
$B$, respectively, for which some of the variances on
single systems is bounded by constants $U_A$, $U_B$ such that
\be
\sum_k\delta^2(\hat{A}_k)\geq U_A  \mbox{ and }\sum_k\delta^2(\hat{B}_k)
\geq U_B.
\ee
Then, we have for separable states
\be
\sum_k\delta^2(\hat{A}_k\otimes\eins+\eins\otimes \hat{B}_k)\geq U_A+U_B
\ee
and violation implies the presence of entanglement.
\end{proposition}

Physically, the LURs state that separable states inherit the uncertainty
relations from their reduced states, which is not the case for entangled
states. Due to this observation
the LURs have attracted a considerable interest, and
a number of interesting
properties have been discovered: LURs can detect bound entangled states
\cite{hofmann2} and can be used to estimate the concurrence
\cite{vicentelur}.
They can be extended to other formulations of the uncertainty principle
\cite{entropic, vicentepollak} and they
can be generalized to nonlocal observables
\cite{oPRL}. Finally, they can be viewed as nonlinear entanglement
witnesses, which improve the CCNR criterion \cite{oPRA}.

For the connection to the CMC we have the following:

\begin{proposition}[Connection to local uncertainty relations]
\label{LURCMC}
A state $\rho$ violates the CMC for symmetric CMs iff it can be
detected by a LUR.
\end{proposition}
{\it Proof:} The proof is given in the Appendix.
$\qed$

This result show that the LURs for appropriate observables
and the CMC are equivalent, however, the CMC
has the major advantage that it can be directly evaluated, while for the
LURs
the appropriate observables have to be identified. Moreover, we can state:

\begin{corol}[Insufficiency of LUR to detect all entangled states]
There exist entangled two qubit states which can not be detected by a LUR,
hence LURs are not a necessary and sufficient criterion for separability.
\end{corol}

{\it Proof:} In the Section \ref{description} we have already constructed a
family of states $\rho_{\varepsilon}$ which cannot be detected by the CMC,
as their symmetric block CM is compatible with a separable as well as an
entangled state. This proves the claim.
$\qed$

\section{The CMC for two qubits}
\label{sec:CMC42q}

After the previous discussion of the situation of
Hilbert spaces
of arbitrary finite dimension, we now turn to the
important simple case of a $2\times 2$-system -- two qubits --
in some more detail. We take
as observables the set
$\{A_k\}=\{B_k\}=\{\eins/\sqrt{2},\sx{}/\sqrt{2},\sy{}/\sqrt{2},
\sz{}/\sqrt{2}\}$
as in Eq.~(\ref{twoqubitloos}).

Since these observables contain the identity, one can easily check
that many terms in the symmetric block CM vanish. Effectively,
$\gamma$ is actually a $6\times 6$ matrix (denoted by
$\gamma^{\text{eff}}$) originating only from the $\{A_k\}$ and
$\{B_k\}$ with $k=1,2,3$ which are not proportional to the identity,
and not by an $8\times 8$ as one could guess from the general
theory.

To characterize the $\kappa_A$ in the CMC,
note that for a pure state $\ket{a}$ on system $A$ we
find, according to
Proposition \ref{CMformsym}, the following properties
of the $4\times 4$ matrix $\gamma(\ketbra{a})$:
\bi
\item[{(i)}] $Rank(\gamma)=2$.
\item[{(ii)}] The nonzero eigenvalues of $\gamma$ are
equal to $1/2$ in a suitable basis.
\ei
We also know that in the chosen basis, the first row as well as the
first column of $\gamma(\ketbra{a})$ vanish, and we have
\be
\gamma(\ketbra{a})=\mathbbm{O}_{1}\oplus
\gamma(\ketbra{a})^{\text{eff}}
\ee
where $\gamma^{\text{eff}}$ is the effective $3 \times 3$ CM as above. This
has to be of rank two with eigenvalues $1/2.$ This implies that
$\gamma(\ketbra{a})^{\text{eff}}$ can be written as
\be
\tilde{\gamma}(\ketbra{a})^{\text{eff}}=\halbe (\eins_3 - \ketbra{\phi_a}
),
\ee
where $\eins_3$ denotes a $3 \times 3$ identity matrix, and
$\ket{\phi_a}\in\R^3$. In fact, any matrix of this form is a
valid CM:

\begin{lem}
For any vector $\ket{\phi}\in \R^3$ a matrix of the form
$(\eins_3 - \ketbra{\phi_a})/2$, is a valid CM of some
two qubit state. Consequently, the set of valid $\kappa_A$
is given by all matrices of the form
\be
\kappa_A  = \halbe (\eins_3-\rho_A)
\label{zweiqubitkappa}
\ee
where $\rho_A$ is a real $3\times 3$ matrix with trace
one and positive eigenvalues.
\end{lem}

{\it Proof:} We have already shown
that the CMs are of the required form, and only
have to argue that any matrix of the form
$X=(\eins_3 - \ketbra{\phi_x})/2$
is a valid CM. To see this, note that unitary transformations of the
$\ket{
a}$ result
in orthogonal transformation on $\gamma(\ketbra{a})^{\text{eff}}$.
Moreover, for the special case of a single qubit any orthogonal
transformation on
$\gamma^{\text{eff}}$ can be generated by a unitary transformation on state
space \cite{HH}, expressing the isomorphism between  the Lie-algebras
$su(2)$ and $so(3)$. Therefore, we can transform $X$ into
$\gamma(\ketbra{a})^{\text{eff}}$
and construct the corresponding state vector $\ket{x}$.
To finish the argument, note that the
set of all $\kappa_A$ is by definition the set of all convex combinations
of pure state CMs.
$\qed$

After having proven this Lemma we can formulate the two-qubit version
of the CMC:

\begin{proposition}[CMC for two qubits]
\label{2qubitcrit}
Let $\rho$ be a state of two qubits and let
\begin{equation}
    \{A_k\}=\{B_k\}=\{\eins/\sqrt{2},\sx{}/\sqrt{2},\sy{}/\sqrt{2},
    \sz{}/\sqrt{2}\}
    \label{equ:2qbasis}
\end{equation}
be the chosen set of observables. Let
$\gamma^\text{eff}$ be the $6\times 6$ CM as mentioned before.
Then the state $\rho$ fulfills the CMC iff there exist $3\times 3$ density
matrices
$\rho_A$ and $\rho_B$ such that
\be
\label{anton}
\gamma^\text{eff} - \halbe\eins_6+\halbe (\rho_A\oplus\rho_B)\geq 0.
\ee
\label{twoqubitCMC}
\end{proposition}
{\it Proof:} The claim follows if we insert the $\kappa$'s
from Eq.\ (\ref{zweiqubitkappa}) into
Proposition \ref{CMcrit}. Note that it suffices to find complex $\rho_A$
and
$\rho_B$. If we can identify such matrices, their real part will
saturate Eq.~(\ref{anton}) as well.
$\qed$

In this form, the problem is a special instance of an efficiently
solvable semidefinite program (SDP) \cite{SDP} in {\em primal} form,
a {\em feasibility} problem .

In general, a SDP consists of a linear function $c^T x$ which is minimized subject
to a semi-definite constraint
\begin{equation}
        F(x)=F_0+\sum_i x_i F_i\ge 0,
\end{equation}
which is linear in the problem variables
$x_i$. Hence the problem is defined by the real vector $c$
and by the hermitian or symmetric matrices $F_i$. If $c=0$,
then the problem is referred to as a {\em feasibility} problem.
Via Lagrange-duality, a {\em dual} problem can be formulated
in which the expression $-\tr(F_0 Z)$ is maximized over a
positive semi-definite (hermitian or symmetric) matrix $Z$,
with the constraints that $\tr(F_i Z)=c_i$.
Since
\begin{equation}
        c^T x+\tr(F_0 Z)=\tr(F(x)Z)\ge 0
\end{equation}
holds true due to the positive semi-definiteness of $F(x)$ and $Z$,
solutions of the dual problem deliver a bound on the solutions of
the primal problem and vice versa, which is referred to as {\em weak
duality}. Finally, if there is a solution to the primal
problem with $F(x)>0$ or a solution to the dual problem with
$Z>0$, then {\em strong} duality holds, meaning that a pair
$(x^*,Z^*)$ exists
such that $c^T x^*+\tr(F_0 Z^*)=0$ holds.
See also Ref.~\cite{Optimization} for an extensive treatment
of the subject.

For the evaluation of the CMC, we can formulate the problem differently,
such that if the primal problem detects the state as entangled, then from
the solution of the dual
problem local operators can be extracted which allow for the detection
of the state with LURs. This is similar in spirit as the solution in
the continuous variable case \cite{sdprog}.

Explicitly, we formulate the primal problem as
\begin{eqnarray}
    \text{min} && -\lambda \label{SDPprimal}\\
    \text{subject\ to} && \gamma^\text{eff}-\kappa_A\oplus\kappa_B \ge 0\nonumber\\
    && \kappa_{A,B}=\frac{1}{2}\big[(1+\lambda)\Eins_3-\rho_{A,B}\big]\ge 0\nonumber\\
    && \tr(\rho_{A,B})=1+\lambda.\nonumber
\end{eqnarray}
In this formulation, the matrices $\kappa_{A,B}$ are positive and have
trace $1+\lambda$. If the constraints can be fulfilled for $\lambda<0$
only, then the state corresponding to $\gamma^\text{eff}$ is
entangled. The SDP can be formulated with block-diagonal matrices
$\{F_i\}$ collecting all the constraints. For instance, by inserting
the definition of $\kappa_{A,B}$ into the first constraint and
expressing the equality constraints by a `$\ge$' and a `$\le$'
constraint, we can write $F_0$ as
\begin{equation}
F_0=(\gamma^\text{eff}-\frac{1}{2}\Eins_6)\oplus
\frac{1}{2}\Eins_3\oplus\frac{1}{2}\Eins_3\oplus(-1)\oplus 1\oplus(-1)\oplus 1,
\end{equation}
and the matrices $F_{i}$ accordingly by choosing a basis for
real, symmetric matrices for the blocks. Without loss of generality,
the matrix $Z$ can be chosen block-diagonal accordingly. In
the order from above we have $Z=Z_1\oplus Z_2^A\oplus Z_2^B\oplus
Z_3^{A1}\oplus Z_3^{A2}
\oplus Z_3^{B1}\oplus Z_3^{B2}$, where $Z_1$ is a $6\times 6$ matrix,
$Z_2^{A,B}$ are of dimension $3\times 3$, and $Z_3^{A,B;1,2}$ are scalar.
The dual problem can then be formulated as
\begin{eqnarray}
    \text{max} && -[\tr(\gamma^\text{eff}Z_1)-1] \label{2qDual}\\
    \text{subject\ to} && -\frac{1}{2}[\tr(Z_1)-\tr(Z_2^A)-\tr(Z_2^B)]\nonumber
    \\
    && \qquad =Z_3^{A1}-Z_3^{A2}+Z_3^{B1}-Z_3^{B2}-1\nonumber\\
    && (Z_1^{A,B})_{i,i}-(Z_2^{A,B})_{i,i}
    =-2(Z_3^{A,B;1}-Z_3^{A,B;2})\nonumber
    \\
    && (Z_1^{A,B})_{i<j}=(Z_2^{A,B})_{i<j},\nonumber
\end{eqnarray}
where $Z_1^{A,B}$ are the single-particle subblocks of system $A$ and $B$,
respectively, and $i$ and $j$ run from $1$ to $3$. It turns out that
$Z_1$ has the properties of an
entanglement witness in the space of covariance
matrices (CM-witness) as in the continuous-variables case  \cite{sdprog}:

\begin{proposition}[CM-Witness from dual program]
\label{2qubitwit}
For every feasible solution $Z$ to the dual problem formulated
above, the matrix $Z_1$ is a CM-witness
in the sense that it fulfills
$\tr(\gamma^\text{eff}_S Z_1)\ge 1$
for all CMs $\gamma^\text{eff}_S$ from separable states.
Hence if $\tr(\gamma^\text{eff}Z_1) < 1$ then the corresponding
state is entangled. Further, it is optimal in the sense
that $\tr(\gamma^\text{eff}Z_1)$
is the minimal value of $\tr(\gamma^\text{eff}X)$ for any $X\ge 0$
of the same dimensions.
\end{proposition}

{\it Proof:}
It follows from weak duality
that $\tr(\gamma^\text{eff} Z_1)\ge 1+\lambda$, hence
$\tr(\gamma^\text{eff}_S Z_1)\ge 1$ holds
for all $\gamma^\text{eff}_S$ from separable states.
In this case, strong duality holds, which we prove by providing
an example:
\begin{equation}
    Z=\frac{3}{2}\Eins_6\oplus\Eins_3\oplus\Eins_3\oplus\frac{3}{4}\oplus 1
    \oplus\frac{3}{4}\oplus 1>0
\end{equation}
fulfills all constraints. Hence there exist $(\lambda^*, Z^*)$
such that $\tr(\gamma^\text{eff}Z_1^*) = 1+\lambda^*$ holds,
and the dual program reaches the minimal value of
$\tr(\gamma^\text{eff} Z_1)$. $\qed$

If the entanglement of a state is detected by a CW-witness
$Z_1$, then it is possible to write down a LUR detecting
the state as well. This is remarkable because it is in general
very difficult to find a LUR detecting the entanglement
of a given state.

\begin{proposition}[LUR observables from witness]
\label{wit2LUR}
Given a CM-witness $Z_1$, it is possible to define
LUR matrices $\{\hat A_k\}$ and $\{\hat B_k\}$ from $Z_1$ such that
\begin{equation}
    \tr(\gamma^\text{eff}Z_1)
    =\sum_k\delta^2(\hat A_k\otimes \Eins+\Eins\otimes \hat B_k)
\end{equation}
holds.
\end{proposition}

{\it Proof:}
The LUR corresponding to $Z_1$ can be extracted as shown
in the proof of Proposition \ref{LURCMC} in the Appendix:
we can spectrally decompose
$Z_1=\sum_k \lambda_k \ketbra{\psi_k}=:
\sum_k\lambda_k\ketbra{\alpha^{(k)}\oplus\beta^{(k)}}$. Defining
the local LUR variables
$\hat{A}_k=\sqrt{\lambda_k}\sum_l\alpha_l^{(k)}A_l$ and
$\hat{B}_k=\sqrt{\lambda_k}\sum_l\beta_l^{(k)}B_l$ we have
for $\rho$ that
$
\tr(Z_1 \gamma^\text{eff})=
\sum_k\delta^2(\hat{A}_k\otimes\eins+\eins\otimes\hat{B}_k),
$
where $\{A_k\}$ and $\{B_k\}$ are defined in Eq.~(\ref{equ:2qbasis}).
$\qed$

\section{Examples}

In this section, we consider bound entangled states of two different
types, and investigate the strength of the different criteria discussed
in this paper.

\begin{figure}[t]
\begin{center}
\includegraphics[width=0.95\columnwidth]{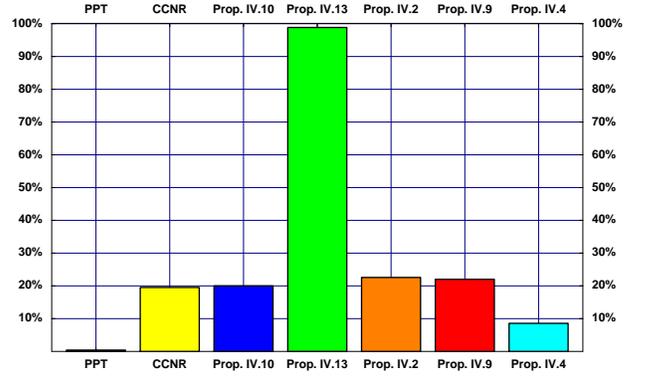}
\caption{(Color online) Detection of $3\times 3$ chessboard states. For the different criteria
the fraction of states which are detected is shown. See text for further details.}
\label{hist}
\end{center}
\end{figure}

In the first example, we take the $3\times 3$ bound entangled states, called
{\it chessboard states}, introduced by D. Bru{\ss} and A. Peres \cite{BP}. They are
defined as
\be
\rho = \mathcal{N} \sum_{j=1}^4 \ketbra{V_j},
\ee
where $\mathcal{N}$ denotes the normalization, and  we used  the unnormalized
vectors
\bear
\ket{V_1}&=&\ket{m,0,ac/n;0,n,0;0,0,0},
\nonumber\\
\ket{V_2}&=&\ket{0,a,0;b,0,c;0,0,0},
\nonumber\\
\ket{V_3}&=&\ket{n,0,0;0,-m,0;ad/m,0,0},
\nonumber\\
\ket{V_4}&=&\ket{0,b,0;-a,0,0;0,d,0}.
\eear
Characterization of the family is done by six real parameters. We
tested all criteria, presented in this paper on randomly generated
chessboard states, where parameters have been drawn from the normal
distribution with zero mean value and standard deviation of two. The
results of this test are presented on the Fig.~\ref{hist}.

As one can see from Fig.~\ref{hist} the most of the states are detected
by bringing first the state in its normal form (Proposition \ref{cmcnfgamsym})
- $98.86 \%$ of all states. The criterion, which uses an estimation of singular
values of the off diagonal block of CM (Proposition \ref{zhangcrit}), which was
also proposed earlier in \cite{china1} detects $22.57 \%$, whereas another
criterion proposed in this paper (Proposition \ref{coptimal}) detects $22.00 \%$.
Moreover the criterion, which uses Schmidt decomposition (Proposition \ref{schmidtdec})
detects $20.00 \%$ which is more or less the same amount as is detected by
CCNR criterion -
$19.52 \%$. Finally the criterion presented in Proposition \ref{vicentecrit}, which
was first proposed by de Vicente \cite{deVic} detects only $8.57 \%$ of randomly
generated chessboard states.

As the second example, we consider $3\times 3$ bound entangled states arising
from an unextendible product basis \cite{UPB}, mixed with white noise:
\bear
\ket{\psi_0}&=&\frac{1}{\sqrt{2}}\ket{0} (\ket{0}-\ket{1}),
\;\;\;
\ket{\psi_1}=\frac{1}{\sqrt{2}}(\ket{0}-\ket{1})\ket{2},
\nonumber\\
\ket{\psi_2}&=&\frac{1}{\sqrt{2}} \ket{2}(\ket{1}-\ket{2}),
\;\;\;
\ket{\psi_3}=\frac{1}{\sqrt{2}}(\ket{1}-\ket{2})\ket{0},
\nonumber
\\
\ket{\psi_4}&=&\frac{1}{3} (\ket{0}+\ket{1}+\ket{2})(\ket{0}+\ket{1}+\ket{2}),
\nonumber \\
\rho_{\rm BE}&=&\frac{1}{4}\Big(\eins-\sum_{i=0}^4\ketbra{\psi_i}\Big),
\nonumber \\
\rho_{\rm UP}(p)&=& p\rho_{\rm BE} + (1-p) \frac{\eins}{9}
\eear
These states are detected by Proposition \ref{cmcnfgamsym} for
$p\geq 0.8723$ while the best known positive map detects them
only for $p\geq 0.8744$ (see \cite{oPRA} and references therein).
Besides this we have also tested all other criteria presented
in this paper. Criteria of Propositions \ref{zhangcrit}, \ref{coptimal}
both detect these states for $p\geq 0.8822$. The criterion derived
for Schmidt decomposed states (Proposition \ref{schmidtdec}) detects
the states for $p\geq 0.8834$, whereas the CCNR criterion detects
them for $p\geq 0.8897$. Finally Proposition \ref{vicentecrit} detects
the states for $p\geq 0.9493$.

Finally, let us shortly comment on the efficiency of the
implementation of all these criteria. The filtering operation
can be implemented quite fast, using the simple algorithm outlined
above takes a few seconds on a desktop computer
($5\times 5$ system:
ca. 6 sec., $10\times 10$ system: ca. 24 sec., $15\times 15$ system:
ca. 72 sec.). Then, the trace norm of $C$ can be quickly computed
as the trace norm of the realignment of the matrix
$\rho-\rho_A\otimes\rho_B$ \cite{china2}. For comparison,
only the first step of the semidefinite program of
Ref.~\cite{doherty} requires already ca. 10 min. for
a $4\times 4$ system, becoming practically unfeasible
for higher dimensions.


\section{Conclusion and outlook}

In this work, we have further developed the ideas of
Ref.~\cite{wir} and investigated the covariance matrix criterion
(CMC). We have shown that this is a strong separability
criterion, which can be simply evaluated. Combined with
filtering it is necessary and sufficient for two qubits and
in  higher dimensions it detects states where the PPT criterion
fails. Moreover, it contains many other separability criteria,
which have been proposed to complement the PPT criterion as
corollaries.

There are several open problems which deserve a further investigation:

\begin{itemize}

\item First, one might study the exact relation
between the CMC for
symmetric CMs and non-symmetric CMs. In the present paper, we
have used only the trace of $\kappa^{A/B}$ for the evaluation, hence this
difference did not become apparent.
However, as the non-symmetric CM
describes the state completely and encodes therefore all information about
the separability properties, this difference might be a way to improve
the CMC. In addition, one could investigate the relation between the linear
part and the nonlinear part of the CM in some more detail. In the proof of
Prop. \ref{uneven} we have seen already that such an investigation may
indeed improve the CMC.

\item Another interesting open question is to relate the CMC to quantitative
statements, such as to the estimation of entanglement measures. One
should expect that the program of Ref.~\cite{Quantitative} linking
violations to quantitative statements on the entanglement content
is applicable to the discussed criteria. Also, similar relations hold for
Gaussian states \cite{giedkecirac}. First steps in this direction
have already been taken in  Refs.~\cite{deVicquant, china1,vicentelur}.

\item Finally, it would be very interesting to develop a theory
similar to ours for entanglement of multiparticle systems. Here,
however, a significant amount of work has yet to be done, as it is not
even obvious how to identify the
object corresponding to the block CM for multipartite
systems.

\end{itemize}

\section{Acknowledgments}
We thank H.J.~Briegel, R. Horodecki and
M.~van~den~Nest for discussions.
This work has been supported by the
EU (OLAQUI, QAP, QICS, SCALA), the FWF,
Microsoft Research, the EPSRC, and the EURYI.

\section{Appendix}

In this appendix, we present the more
technical proofs of our previous statements.

{\it Proof of Proposition \ref{OUconnection}:} Let us first explain some
properties of the matrix
$\Gamma$. This matrix has entries which are just the basis
vectors $G_i$ written as columns. Moreover,
$\Gamma\Gamma^{\dagger}=\Eins = \Gamma^{\dagger}\Gamma$,
i.e., $\Gamma$ is a unitary, since
\begin{eqnarray}
(\Gamma^{\dagger}\Gamma)_{i,j}
&=&
\sum_k\Gamma^{\dagger}_{ik}\Gamma_{kj}
=
\sum_{\alpha,\beta}(G^{\alpha,\beta}_i)^* G^{\alpha,\beta}_j
\nonumber\\
&=&
\sum_{\alpha,\beta}(G^{\beta, \alpha}_i) G^{\alpha,\beta}_j=
\tr(G_iG_j)=\delta_{i,j},
\end{eqnarray}
where we have used the orthogonality and hermiticity of $G_i.$
However $\Gamma$ is a special unitary, since the columns correspond
to orthonormal Hermitian observables.
Now we have in Eq.~(\ref{generatedO})
\be
\sum_j O_{i,j}G_j^{\alpha,\beta} =
\sum_j G^{\alpha,\beta}_j\left(O^T\right)_{j,i}=\left(\Gamma O^T\right)_{
\alpha,\beta|i},
\label{onehand}
\ee
where we have used the definition of $\Gamma$ and
the fact that the expression in the middle of
Eq.~(\ref{onehand}) is nothing but $i$-th column
of $\Gamma O^T$. Conversely,
\begin{align}
(U & G_{i}U^{\dagger})_{\alpha,\beta}=U_{\alpha,\delta}
G^{\delta,\gamma}_{i}
U^{\dagger}_{\gamma,\beta}=U_{\alpha,\delta}U^{\ast}_{\beta,\gamma}\Gamma
_{\delta,\gamma |i}
\nonumber
\\
&=\left(U\otimes U^{\ast}\right)_{\alpha,\beta,\delta,\gamma}
\Gamma_{\delta,\gamma |i}
=\left(U\otimes U^{\ast}\Gamma\right)_{\alpha,\beta | i},
\end{align}
where we have used the definition of $\Gamma$ and that
$A_{i,k}\otimes B_{l,m}\equiv (A\otimes B)_{i,l,k,m}$.
Therefore we can write
\begin{align}
O^T&=\Gamma^{\dagger}(U\otimes U^{\ast})\Gamma=
\Gamma^T(U^{\ast}\otimes U)\Gamma^{\ast},
\nonumber
\\
O&=\Gamma^{\dagger}(U^{\dagger}\otimes U^T)\Gamma=\Gamma^T(U^T\otimes
U^{\dagger})\Gamma^{\ast},
\end{align}
where we used that $O$ is real.
With these representations, we can finally check
the orthogonality of the $O$ as
\begin{eqnarray}
O^TO &=&
\Gamma^T(U^{\ast}\otimes U)\Gamma^{\ast}\Gamma^T(U^T\otimes U^{\dagger})\Gamma^{\ast}
\nonumber\\
&=&\Gamma^T(U^{\ast}\otimes U)\Eins(U^T\otimes U^{\dagger})\Gamma^{\ast}
\nonumber\\
&=&\Gamma^T(U^{\ast}U^T\otimes UU^{\dagger})\Gamma^{\ast}
=\Gamma^T\Gamma^{\ast}=\Eins.
\end{eqnarray}
$\qed$

{\it Proof of Proposition \ref{symm-nonsymm}:} First note that if we write
$\gamma$
as in Eq.~(\ref{BCM}) in the blockwise form,
$A,B$ correspond to CMs of the subsystems $A,B$
and $C$ has entries of the form $\mean{A_i\otimes B_j}-\mean{A_i}\mean{B_j}
$, where $A_i,B_j$ are observables taken for subsystems $A,B$.

The condition $\gamma=\gamma^{S}$ is equivalent to the condition
$\gamma=\gamma^T$, in particular $A=A^T$ and $B=B^T$. If we change
the local bases on $A$ and $B$ via $O=O^A\oplus O^B$
the CM gets to
\be
\gamma^{\prime}=(O^A\oplus O^B)\gamma (O^A\oplus O^B)^T.
\label{bastranseq}
\ee
As we can immediately see $\gamma^{\prime T}=(O^A\oplus O^B)\gamma^T (O^A
\oplus O^B)^T=
\gamma^{\prime}$ if and only if $\gamma^T=\gamma$,
so the symmetry of CM
does not depend on the particular choice of basis in observable space.

Therefore we are able to choose the standard basis. Let us consider only
 subsystem $A$, i.e., left upper block of matrix $\gamma$, and let us
 assume
that $A=A^T$ holds. As we have showed already, we can obtain $\rho_A$
from the matrix $A$ by use of the commutators $A_{i,j}-A_{j,i}=
\mean{\Kommut{M_i^A}{M_j^A}}$.
However, all those commutators vanish for the case $A=A^T$. Since then
$\mean{X_{k,l}}=\mean{Y_{k,l}}=0$ for all $ k,l$, it follows that
$\rho_A$
is diagonal. The diagonal elements can be also determined as in Prop.\
\ref{cm-dm} and since also $\mean{Z_{k,l}}=0 $ for all $ k,l$, it
follows that $\rho_{A}=\eins/d_A$, which completes the first part of the
proof.

Finally, local unitary transformations are only a subclass of the
orthogonal transformations considered before, hence
$\gamma=\gamma^T$ cannot be achieved by a local unitary
transformation of $\rho$ neither.
$\qed$

{\it Alternative proof of Proposition \ref{BCMUIN} for Ky-Fan norms:}
For a matrix as in Eq.~(\ref{claim}) the following condition
has to be fulfilled:
\be
\begin{pmatrix}\bra{\alpha}\\ \bra{\beta}\end{pmatrix}
\begin{pmatrix}A&C\\ C^T&B\end{pmatrix}
\begin{pmatrix}\ket{\alpha}& \ket{\beta}\end{pmatrix}
\geq 0,
\ee
for all vectors $\ket{\alpha}, \ket{\beta}$, which implies that
$
\bra{\alpha}A\ket{\alpha} + \bra{\beta}B\ket{\beta} \geq
2\bra{\alpha}C\ket{\beta},
$
where we took $-\ket{\beta}$ instead of $\ket{\beta}$ for
convenience. Especially, we can take $\ket{\alpha}=\alpha\ket{\psi_k}$
and $\ket{\beta}=\beta\ket{\phi_k}$, where the vectors $\ket{\psi_k}$
and $\ket{\phi_k}$ are singular vectors from the singular value
decomposition of $C$ and $\bra{\psi_k}C\ket{\phi_k}=\sigma_k(C)$
is the  $k$-th singular value. Hence
\be
\alpha^2\bra{\psi_k}A\ket{\psi_k} + \beta^2\bra{\phi_k}B\ket{\phi_k} \geq
2\alpha\beta\bra{\psi_k}C\ket{\phi_k}.
\ee
Note that $\bra{\psi_k}A\ket{\psi_k}$ and $\bra{\psi_k}A\ket{\psi_k}$ are
greater than zero, because $A$ and $B$ are positive semi-definite matrices.
Taking the sum over $k$ and noting that for $A$ and $B$ expressions like
$\sum_{k=1}^K \bra{\psi_k} A \ket{\psi_k}$ are a lower bound on the $K$-th
Ky-Fan norm \cite{HornJohnson} we get
\be
\alpha^2 \Vert A \Vert_{KF}+\beta^2 \Vert B \Vert_{KF}
\geq 2\alpha\beta \Vert C \Vert_{KF}.
\label{interestingequation}
\ee
The last formula is necessary and sufficient condition for
the $2\times 2$
matrix
\be
\begin{pmatrix}\Vert A \Vert_{KF}& \Vert C \Vert_{KF}\\
\Vert C \Vert_{KF}& \Vert B \Vert_{KF}\end{pmatrix} \geq 0
\ee
to be positive semi-definite and having a non-negative determinant, from
which the claim follows.
$\qed$

{\it Proof of Proposition \ref{LURCMC}:} The proof is an adaption of a
similar proof given in Ref.~\cite{oPRL}. We will often use the property that
CMs can be used to compute variances. Imagine $N=\sum_k \nu_k M_k$
is a linear combination of the $M_k$ with $\nu_k\in\R$, then
\be
\delta^2(N)=\sum_{k,l}\nu_k\nu_l(\mean{M_kM_l}-\mean{M_k}\mean{M_l})\\
=\bra{\nu}\gamma(\{M\})\ket{\nu}.
\ee

Let us now assume that $\rho$ violates
the LURs and we can find $\hat{A}_k$, $\hat{B}_k$, $U_A$ and $U_B$
as in Proposition \ref{LURs}. We assume that the CMC is fulfilled,
i.e., there exist $\kappa_A$ and $\kappa_B$ such that for the CM
$\gamma$ we have $\gamma\geq \kappa_A\oplus\kappa_B$. We can write
\be
\hat{A}_k =\sum_l\alpha_l^{(k)}A_l
\mbox{ and }
\hat{B}_k =\sum_l\beta_l^{(k)}B_l
\ee
where the $\{A_k\}$ and $\{B_k\}$
are the observables chosen  in the definition of
$\gamma$. This leads to
$\delta^2(\hat{A}_k\otimes\eins+\eins\otimes\hat{B}_k)=
\bra{\alpha^{(k)}\oplus\beta^{(k)}}\gamma\ket{\alpha^{(k)}\oplus\beta^{(k)}
}.$
Also, by definition
\be
\kappa_A\oplus\kappa_B=\sum_l p_l\gamma(\ketbra{a_l})\oplus\gamma(\ketbra
{b_l})
\ee
and hence
$
\bra{\alpha^{(k)}\oplus\beta^{(k)}}\kappa_A\oplus\kappa_B\ket{\alpha^{(k)}
\oplus\beta^{(k)}}
=\sum_l p_l [\delta^2(\hat{A}_k)_{\ketbra{a_l}}+\delta^2(\hat{B}_k)_{
\ketbra{b_l}}].
$
But then summing over $k$ yields
\begin{align}
&\sum_k\delta^2(\hat{A}_k\otimes\eins+\eins\otimes\hat{B}_k)
\\
&\geq
\sum_{k,l} p_l
\big[
\delta^2(\hat{A}_k)_{\ketbra{a_l}}+\delta^2(\hat{B}_k)_{\ketbra{b_l}}
\big]
\nonumber
\\
&
\geq
\min_{\ketbra{a}} \sum_k \big[\delta^2(\hat{A}_k)_{\ketbra{a}}\big]+
\min_{\ketbra{b}} \sum_k \big[\delta^2(\hat{B}_k)_{\ketbra{b}}\big]
\nonumber
\\
&
\geq U_A + U_B,
\nonumber
\end{align}
which is a contradiction to our assumption that $\rho$ violates
the LURs.

To show the converse
direction, let us assume that $\vr$ violates the CMC.
Let us define a set of matrices as
$X=\{x | x=\kappa_A\oplus\kappa_B+P \mbox{ with } P\geq 0\}$,
which geometrically is a closed convex cone. Using this definition,
we can formulate the CMC differently, by saying that {\it if $\rho$ is
separable, then $\gamma\in X$}. As our $\rho$
violates the CMC, we have $\gamma\notin X$.

According to a corollary to the Hahn-Banach theorem \cite{hbt} for
each $\gamma\notin X$ there exist a symmetric matrix  $W$ and a number
$C$ such that $\tr(W\gamma)< C$ while
\begin{equation}
    \tr(Wx)>C\,\forall x\in X.
    \label{witness}
\end{equation}
Since $X$ is a non-compact cone, and we can add arbitrary positive
operators to the elements of $X,$ we can conclude that
$\tr(WP)\geq 0$ has to hold for all $P\geq 0$, and consequently we
have $W \geq 0$. Now let us make use of spectral decomposition of
$W$ and write  $W=\sum_k \lambda_k \ketbra{\psi_k}=:
\sum_k\lambda_k\ketbra{\alpha^{(k)}\oplus\beta^{(k)}}$. Defining
$\hat{A}_k=\sqrt{\lambda_k}\sum_l\alpha_l^{(k)}A_l$ and
$\hat{B}_k=\sqrt{\lambda_k}\sum_l\beta_l^{(k)}B_l$ we have
for $\rho$ that
\be
\tr(W\gamma)=
\sum_k\delta^2(\hat{A}_k\otimes\eins+\eins\otimes\hat{B}_k).
\label{W2LUR}
\ee
Furthermore, by definition we have for all $\kappa_A\oplus\kappa_B\in X$
and from Proposition \ref{gammaprop} it follows that all
$\gamma_A\oplus\gamma_B\in X$. Hence for each product
state $\rho=\rho_A\otimes\rho_B$ we have
$
C<\tr(W\gamma_A\oplus\gamma_B)=
\sum_k[\delta^2(\hat{A}_k)_{\rho_A}+\delta^2(\hat{B}_k)_{\rho_B}].
$
This implies that
\begin{align}
C&<
\min_{\rho_A,\rho_B}
\big[\sum_k(\delta^2(\hat{A}_k)_{\rho_A}+\delta^2(\hat{B}_k)_{\rho_B})\big]
\nonumber
\\
&<\min_{\rho_A}
\big[\sum_k\delta^2(\hat{A}_k)_{\rho_A}\big]+
\min_{\rho_B}
\big[\sum_k \delta^2(\hat{B}_k)_{\rho_B}\big]
\nonumber
\\
&=:U_A + U_B
\end{align}
Finally, since the CMC is violated, $\gamma\notin X$ and
$
\sum_k\delta^2(\hat{A}_k\otimes\eins+\eins\otimes\hat{B}_k)=\tr(W\gamma)<
C<U_A + U_B$
leading to a violation of the LURs criterion.

Note that in principle this proof also applies to the CMC for non-symmetric
CMs.
Then, however, the "observables" in the LURs will be non-hermitian, their
variance has to be defined as
$\delta^2(X)=\mean{X X^\dagger}-\mean{X}\mean{X^\dagger}$ and their
physical interpretation is not so clear.
$\qed$


\begin{thebibliography}{99}

\bibitem{hororeview}
    R.\ Horodecki, P.\ Horodecki, M.\ Horodecki, and
    K.\ Horodecki,
    quant-ph/0702225.

\bibitem{morereviews}
    D.\ Bru{\ss}, J.I.\ Cirac, P.\ Horodecki, F.\ Hulpke,
    B.\ Kraus, M.\ Lewenstein, and A.\ Sanpera,
    J.\ Mod.\ Opt.\ {\bf 49}, 1399 (2002);
    B.\ M.\ Terhal,
    J.\ Theo.\ Comp.\ Sc.\ {\bf 287}(1), 313
    (2002);
    M.B.\ Plenio and S.\ Virmani,
    Quant.\ Inf.\ Comp.\ {\bf 7} , 1 (2007).

\bibitem{Gurvits}
    L.\ Gurvits, quant-ph/0303055;
    L.M.\ Ioannou, Quant.\ Inf.\ Comp.\ {\bf 7}, 335 (2007).

\bibitem{ppt1}
        A.\ Peres,
        Phys.\ Rev.\ Lett \ {\bf 97}, 1413 (1996).

\bibitem{ppt2}
        M.\ Horodecki,\ P.\ Horodecki, and R.\ Horodecki,
        Phys.\  Lett.\  A \ {\bf 223}, 1 (1996).

\bibitem{horobound}
    P.\ Horodecki,
    Phys.\ Lett.\ A {\bf 232}, 333 (1997).

\bibitem{positivemaps}
    B.\ M.\ Terhal,
    Lin.\ Alg.\ Appl.\ {\bf 323}, 61 (2000);
    M.\ Piani,
    Phys.\ Rev.\ A {\bf 73}, 012345 (2006);
    H.-P.\ Breuer,
    Phys.\ Rev.\ Lett.\ {\bf 97}, 080501 (2006);
    D.\ Chruscinski and  A.\ Kossakowski,
    quant-ph/0606211.


\bibitem{redcrit}
        M.\ Horodecki and P.\ Horodecki,
        Phys.\ Rev.\ A  {\bf 59}, 4206 (1999).

\bibitem{kempenielsen}
        M.\ A.\ Nielsen and J.\ Kempe,
        Phys.\ Rev.\ Lett.\ {\bf 86}, 5184 (2001).

\bibitem{entropies}
    R.\ Horodecki, P.\ Horodecki, and M.\ Horodecki,
    Phys.\ Lett.\ A
    {\bf 210}, 377, (1996);
    N.\ J.\ Cerf and C.\ Adami,
    Phys.\ Rev.\ Lett.\ {\bf 79}, 5194 (1997);
    S.\ Abe and A.K.\ Rajagopal,
    Physica A {\bf 289}, 157 (2001).

\bibitem{vollbrecht}
    K.\ G.\ H.\ Vollbrecht  and M.\ M.\ Wolf,
    J.\ Math.\ Phys.\ {\bf 43}, 4299 (2002).

\bibitem{nagasaki}
    T.\ Hiroshima,
    Phys.\ Rev.\ Lett.\ {\bf 91}, 057902 (2003).

\bibitem{doherty}
     A.\ C.\ Doherty, P.\ A.\ Parrilo, and  F.\ M.\ Spedalieri,
         Phys.\ Rev.\ Lett.\ {\bf 88}, 187904 (2002).

\bibitem{ourold}
      J.\ Eisert, P.\ Hyllus, O.\ G{\"u}hne, and M.\ Curty,
      Phys.\ Rev.\ A {\bf 70}, 062317 (2004).

\bibitem{ccncrit}
        O.\ Rudolph,
        quant-ph/0202121;
        K.\ Chen and L.-A.\ Wu,
        Quant.\ Inf.\ Comp.\ {\bf 3}, 193 (2003);
        M.\ Horodecki, P.\ Horodecki, and R.\ Horodecki,
        Open Syst.\ Inf.\ Dyn.\ {\bf 13}, 103 (2006).

\bibitem{deVic}
        J.I.\ de\ Vicente,
        Quantum\ Inf.\ Comput.\ {\bf 7}, 624 (2007).

\bibitem{deVicquant}
        J.I.\ de\ Vicente,
        J.\ Phys.\ A {\bf  41}, 065309 (2008).

\bibitem{LUR}
        H.F.\ Hofmann and S.\ Takeuchi,
        Phys.\ Rev.\ A\ {\bf 68}, 032103 (2003).

\bibitem{YuLiu}
        S.\ Yu and N.\ Liu,
        Phys.\  Rev.\  Lett.\  {\bf 95}, 150504 (2005).

\bibitem{china1}
    C.-J.\ Zhang, Y.-S.\ Zhang, S.\ Zhang, and
    G.-C.\ Guo,
    Phys.\ Rev.\ A\ {\bf 76}, 012334 (2007).

\bibitem{china2}
    C.-J.\ Zhang, Y.-S.\ Zhang, S.\ Zhang, and
    G.-C. Guo,
    Phys.\ Rev.\ A\ {\bf 77}, 060301(R) (2008).

\bibitem{wir}
    O.\ G\"uhne, P.\ Hyllus, O.\ Gittsovich, and J.\ Eisert,
    Phys.\ Rev.\ Lett.\ {\bf 99}, 130504 (2007).



\bibitem{mira}
    A.\ Miranowicz, M.\ Piani, P.\ Horodecki, and
    R.\ Horodecki, quant-ph/0605001.

\bibitem{robertson}
        H.P.\ Robertson,
        Phys.\ Rev.\ {\bf 46}, 794 (1934).

\bibitem{Vourdas}
    A.\ Vourdas, Rep.\ Prog.\ Phys.\ {\bf 67},
    267 (2004).

\bibitem{Weyl}
    Let $X(q)|j\rangle = |j+q\rangle$
    and $Z(p)|j\rangle = e^{2\pi i pj /d}|j\rangle$ be shift
    and multiply operators,
    then the Weyl operators are defined
    as
    \begin{equation}
        W(q,p)=e^{\pi i (d+1) p q /d} Z(p) X(q).
    \end{equation}


\bibitem{Rud2}
        O.\ Rudolph,
        Phys.\ Rev.\ A\ {\bf 67}, 032312 (2003).

\bibitem{blockremark} We denote by $A\oplus B$ a $2 \times 2$ block matrix with $A$ and $B$
on the diagonal, and zero matrices elsewhere.

\bibitem{oPRL}
        O.\ G\"uhne,
        Phys.\ Rev.\ Lett.\ {\bf 92}, 117903 (2004).

\bibitem{Survey}
    J.\ Eisert and M.B.\ Plenio,
    Int.\ J.\ Quant.\ Inf.\ {\bf 1}, 479 (2003).

\bibitem{Survey2}
    P.\ van Loock and S.\ Braunstein, Rev.\ Mod.\ Phys.\
    {\bf 77}, 513 (2005).

\bibitem{CVbe}
    R.F.\ Werner and M.M.\ Wolf,
    Phys.\ Rev.\ Lett.\ {\bf 86}, 3658 (2001).

\bibitem{SDP}
    L.\ Vandenberghe and S.\ Boyd, SIAM Rev.\ {\bf 38}, 49 (1996);
    C.\ Helmberg, Eur.\ J.\ Oper.\ Res.\ {\bf 137}, 461 (2002).

\bibitem{sdprog}
        P.\ Hyllus and J.\ Eisert,
        New\ J.\ Phys.\ {\bf 8}, 51 (2006).


\bibitem{HornJohnson}
        R.\ A.\ Horn and  C.\ R.\ Johnson,\
        {\it Topics in Matrix Analysis}
        (Cambridge  University  Press, Cambrigde, 1991).

\bibitem{WE}
    Let $A,B$ be Hermitian $n\times n$-matrices.
    Then,
    the non-increasingly ordered eigenvalues
    satisfy
    \begin{eqnarray}
        \lambda_j
        (A+B) \leq
        \lambda_i
        (A) + \lambda_{j-i+1}
        (B), i\leq j,\\
        \lambda_j
        (A+B) \geq \lambda_i
        (A) + \lambda_{j-i+n}
        (B), i\geq j.
    \end{eqnarray}

\bibitem{oPRA}
    O.\ G\"uhne, M.\ Mechler, G.\ T\'oth, and  P.\ Adam,
    Phys.\ Rev.\ A\ {\bf 74}, 010301(R) (2006).

\bibitem{norway}
        J.M.\ Leinaas,\ J.\ Myrheim, and E.\ Ovrum,
        Phys.\ Rev.\ A\ {\bf 74}, 012313 (2006).

\bibitem{frankfilterqubit}
    F.\ Verstraete,\ J.\ Dehaene,
    and B.\ De\ Moor,
        Phys.\ Rev.\ A\ {\bf 64}, 010101 (R) (2001).

\bibitem{frankfiltermulti}
    F.\ Verstraete,\ J.\ Dehaene, and B.\ De\ Moor,
        Phys.\ Rev.\ A\ {\bf 68}, 012103 (2003).

\bibitem{filtop}
        A.\ Kent,\ N.\ Linden and S.\ Massar,
        Phys.\ Rev.\ Lett.\ {\bf 83}, 2656 (1999);
        F.\ Verstraete,\ J.\ Dehaene, and B.\ De\ Moor,
        Phys.\ Rev.\ A\ {\bf 65}, 032308 (2002).

\bibitem{hofmann2}
    H.F.\ Hofmann,
    Phys.\ Rev.\ A {\bf 68}, 034307 (2003).

\bibitem{vicentelur}
    J.I.\ de\ Vicente,
    Phys.\ Rev.\ A\ {\bf 75}, 052320 (2007).

\bibitem{entropic}
    O.\ G\"uhne and M.\ Lewenstein,
    Phys.\ Rev.\ A {\bf 70}, 022316 (2004).

\bibitem{vicentepollak}
    J.I.\ de Vicente and  J.\ S\'anchez-Ruiz,
    Phys.\ Rev.\ A {\bf 71}, 052325 (2005).

\bibitem{HH}
        R.\ Horodecki and M.\ Horodecki,
        Phys.\ Rev.\ A\ {\bf 54}, 1838 (1996).

\bibitem{Optimization}
    S.\ Boyd and L.\ Vandenberghe, {\em Convex Optimization}
    (Cambridge University Press, Cambridge, 2004).

\bibitem{BP}
    D.\ Bru{\ss} and  A.\ Peres,
    Phys.\ Rev.\ A {\bf 61}, 30301(R) (2000).

\bibitem{UPB}
    C.H.\ Bennett, D.P.\ DiVincenzo, T.\ Mor, P.W.\ Shor,
    J.A.\ Smolin, and B.M.\ Terhal,
    Phys.\ Rev.\ Lett.\ {\bf 82} 5385 (1999).

\bibitem{hbt}
        See,\ e.g.\ F.\ Hirzebruch and W.\ Scharlau,
        {\it Einf\"uhrung in die Funktionalanalysis}
        (BI, Mannheim, 1971).

\bibitem{Quantitative}
        K. Chen, S. Albeverio, S.-M. Fei,
        Phys. Rev. Lett. {\bf 95}, 040504 (2005);
        K. Chen, S. Albeverio, S.-M. Fei,
        Phys. Rev. Lett. {\bf 95}, 210501 (2005);
        H.-P.\ Breuer,
    J.\ Phys.\ A {\bf 39}, 11847 (2006);
        K.M.R.\ Audenaert and M.B.\ Plenio,
    New J.\ Phys.\ {\bf 8}, 266 (2006);
    J.\ Eisert, F.G.S.L.\ Brandao, and K.M.R.\ Audenaert,
    New J.\ Phys.\ {\bf 9}, 46 (2007);
    O.\ G{\"u}hne, M.\ Reimpell, and R.F.\ Werner,
    Phys.\ Rev.\ Lett.\ {\bf 98}, 110502 (2007);
        F. Mintert, Phys. Rev. A {\bf 75}, 052302 (2007);
        O.\ G{\"u}hne, M.\ Reimpell, and R.F.\ Werner,
        arXiv:0802.1734.

\bibitem{giedkecirac}
G. Giedke and J.I. Cirac,
Phys. Rev. A {\bf 66}, 032316 (2002).

\end{thebibliography}
\end{document}